\documentstyle[prc,aps,psfig]{revtex}
\begin{document}
\draft
\title{Effects of hyperons on the dynamical deconfinement transition 
\\in cold neutron star matter}
\author{Kei Iida$^{1}$ and Katsuhiko Sato$^{1,2}$}
\address{$^{1}$Department of Physics, University of Tokyo,
7-3-1 Hongo, Bunkyo, Tokyo 113-0033, Japan}
\address{$^{2}$Research Center for the Early Universe, 
University of Tokyo,
7-3-1 Hongo, Bunkyo, Tokyo 113-0033, Japan}
\date{\today}
\maketitle

\begin{abstract}
The influence of the presence of hyperons in dense hadronic matter on 
the quantum nucleation of quark matter is examined at low temperatures 
relevant to neutron star cores.  We calculate the equation of state
and the composition of matter before and after deconfinement by using 
a relativistic mean-field theory and an MIT bag model, respectively;
the case in which hyperons are present in the hadronic system is 
considered, together with the case of the system without hyperons.
We find that
strangeness contained in hyperons acts to reduce a density jump at 
deconfinement as well as a lepton fraction in the hadronic phase. 
%considerably.  
As a result of these reductions, a quark matter droplet 
being in a virtual or real state has its effective mass 
%drastically 
lightened and its electric charge diminished into nearly zero.  The 
Coulomb screening of leptons on the droplet charge, which has 
significance to the droplet growth after nucleation in the absence of
hyperons, is thus shown to be of little consequence.  If the effective
droplet mass is small enough to become comparable to the height of the
potential barrier, the effect of relativity brings about an
exponential increase in the rate of droplet formation via quantum
tunneling, whereas the role played by energy dissipation in
decelerating the droplet formation,
%the formation rate, 
dominant for matter without hyperons,
becomes of less importance.  Independently of the presence of
hyperons, the dynamical compressibility of the hadronic phase is 
unlikely to affect the quantum nucleation of quark matter at
temperatures found in neutron star interiors.  For matter with and
without hyperons, we estimate the overpressure needed to form the 
first droplet in the star during the compression due to stellar 
spin-down or mass accretion from a companion star.  The temperature at
which a crossover from the quantum nucleation to the Arrhenius-type 
thermal nucleation takes place is shown to be large compared with the 
temperature of matter in the core.  We also determine the range of the
bag-model parameters such as the bag constant, the QCD fine structure
constant, and the strange quark mass where quark matter is expected
to occur in the star.  
\end{abstract}
\pacs{PACS number(s): 26.60.+c, 12.38.Mh, 64.60.Qb, 97.60.Jd}

% body of paper here

%\begin{center}
%{\bf I. INTRODUCTION}
%\end{center}
%\vspace*{0.5cm}
\section{INTRODUCTION}
\label{sec:intro}

A neutron star core, at densities near and just above the normal 
nuclear density $n_{0}\approx 0.16$ fm$^{-3}$, consists of uniform 
nuclear matter being electrically neutral and roughly in $\beta$ 
equilibrium between its components such as neutrons, protons, 
electrons, and muons.  For still higher densities, it is possible that 
various forms of matter including hyperons, meson (pion or kaon) 
condensates, quark matter droplets, and/or quark-gluon plasmas might 
become energetically favorable; their presence is predicted to play an 
important role in determining the structure and evolution of neutron 
stars [1].  Basically, these forms are thought to occur via 
pressure-induced phase transitions, whose nature, both static and 
dynamical, is still uncertain in the absence of adequate information 
about the properties of matter in the core such as the mutual 
interactions between its possible constituents and the equation of 
state.

The possible existence of quark matter in neutron stars started to be 
considered about two decades ago mainly by comparing the energies of
hadronic and quark matter at zero temperature [2].  Using a 
separate physical description of the two phases, the transition from 
uniform hadronic matter to uniform quark matter, which proceeds 
suddenly at constant pressure, was expected to occur in the density 
range $\sim 5$--$10 n_{0}$.  The presence of uniform quark matter 
would enhance neutrino luminosity and neutron star cooling via quark 
Urca processes [3] as well as soften the equation of state effectively 
[4].  However, the transition in the $\beta$-equilibrated matter, 
which contains two conserved charges, i.e., electric charge and baryon
number, should proceed through a mixed phase persisting over a finite
range of pressure, as predicted by Glendenning [5] according to Gibbs'
criteria for phase equilibrium.  In the low pressure regime of the
mixed phase, as asserted by Glendenning [5] and Heiselberg $et$ $al$.\
[6], quark matter droplets form a Coulomb lattice embedded in a sea of
hadrons and in a roughly uniform sea of electrons and muons.  This
spatial structure can arise because the presence of strange and down
quarks in the negatively charged quark phase plays a role in decreasing 
the electron and muon Fermi energies and in increasing the proton 
fraction in the positively charged hadronic phase.  With 
increasing pressure, the shape of quark matter changes from spheres to
rods and then to plates, until the part of quark matter and that
of hadronic matter begin to be replaced by each other.  After further 
changes of the shape of hadronic matter from plates to rods and then
to spheres, the system turns into uniform quark matter at the highest 
pressure.  Such
%The ordered 
structure of one- to three-dimensional periodicity, which has a
characteristic length $\sim10$ fm,
%various dimensions occurring in the mixed phase 
would contribute not only to quake phenomena, possibly
relevant to pulsar glitches, via its elastic properties [7], but also
to enhanced neutron star cooling due to opening of the quark phase
space for neutrino-generating processes that are inhibited in a 
translationally invariant system [8].  This sequence of geometrical
structure 
%in the mixed phase 
comes from the competition between the
surface and Coulomb energies, indicating that these structural 
transitions are of first order.
%resulting immediately in first-order 
%nature of these structural transitions.  
The mechanism of quark-hadron 
phase transitions at nonzero densities, however, has not yet been 
established by full QCD calculations, e.g., by Monte Carlo 
calculations of lattice gauge theory.

Kinetics of the quark-hadron phase transitions has been studied 
primarily in the context of the early universe and ultrarelativistic 
collisions of heavy ions [9].  The astrophysical situation in which 
such kinetics could be significant is compression of dense stellar
matter during collapse of a massive star's core, spin-down of a
neutron star, or mass accretion onto a neutron star 
from a companion star.
%in a neutron star due to stellar collapse, stellar spin-down, or mass 
%accretion from a companion star.  
The deconfinement transition in 
hadronic matter not containing strangeness, if proceeding via 
homogeneous nucleation [10], begins with appearance of a critical-size
droplet of two-flavor (up and down) quark matter induced by
fluctuations.  Since a precritical droplet moves back and forth on a 
time scale for strong interactions, $\sim 10^{-23}$ s, which is many
orders of magnitude smaller than that for weak interactions, flavor must
be conserved during the precritical situation.  As a consequence, a 
critical droplet of three-flavor quark matter containing strange
quarks is unlikely to occur in spite of its stability over two-flavor
quark matter.  It is expected that a critical droplet of up and down
quark matter 
%may 
appears either via thermal activation of the energy
barrier separating the initial metastable (hadronic) phase from the
stable (quark) phase in configuration space at high temperatures 
typical of matter created in highly energetic heavy-ion collisions and
in stellar collapse, or via quantum penetration of this barrier at low
temperatures appropriate to neutron star cores.  We also note that
during the growth of the critical droplet, weak processes producing
strange quarks proceed fully in the stellar interiors in contrast to
the case of heavy-ion collisions.

In our recent papers [11,12], the time required to form a two-flavor
quark matter droplet in a neutron star core, originally composed of
nuclear matter in $\beta$ equilibrium, was calculated for pressures
near the point of the static deconfinement transition by using a
quantum tunneling analysis incorporating the electrostatic energy and
the effects of energy dissipation. It was found that the dissipation
effects, governed by collisions of low-energy excitations in the
hadronic phase with the droplet surface, significantly increase
%mainly determine 
the degree
of overpressure needed to form a droplet in the star, and that the
formed droplet develops into bulk matter due to the screening of
leptons on the droplet charge.  
%This result implies that the mixed phase consisting of quark 
%and nuclear matter is unlikely to occur in the star.  
We also discovered that the nucleation is likely to proceed
via quantum tunneling rather than via thermal activation at
temperatures below $\sim 0.1$ MeV typical of neutron star matter.  It
is of interest to note that strangeness may 
%possibly 
be present in the
hadronic phase in the form of hyperons ($\Lambda$, $\Sigma^{\pm}$,
$\Sigma^{0}$, etc.) [13,14] or in the form of a kaon condensate [15].
The presence of strange quarks in hyperons or kaons, 
%may well 
altering not only the equation of state and the composition before 
and after deconfinement but also the quark-hadron interfacial energy, 
may affect the nature of the formation and growth of a quark matter 
droplet.

A new physics content that we consider in this paper is the influence
of the presence of hyperons on the dynamical deconfinement transition 
that may occur in the hadronic core of a neutron star accreting matter
from a companion or rotating down.  We particularly evaluate at what
pressure a quark matter droplet forms via quantum tunneling in the 
metastable phase of hadronic matter containing hyperons (hereafter 
referred to as hyperonic matter).  For this purpose, the thermodynamic
properties of hyperonic matter such as the equation of state and the
composition are estimated from a relativistic mean-field theory; 
we use the one 
with the parameters incorporating a binding energy of $\Lambda$ in
saturated nuclear matter [16].  By using an MIT bag model for the 
thermodynamic properties of quark matter, we calculate not only the 
pressure at which a deconfinement transition from hyperonic matter in 
$\beta$ equilibrium takes place statically subject to flavor
conservation, but also the corresponding jump of baryon density.  In
the overpressure regime, the time needed to form a quark matter
droplet is obtained using a theory of quantum nucleation developed by
Lifshitz and Kagan [17]; we build into this theory the effects of
Coulomb energy, special relativity, energy dissipation [18], and
dynamical [19] and static compressibility.  Some of these effects are
found to be crucial
%essential 
for the formation and growth of a quark matter 
droplet.  We then estimate the critical overpressure required to form
the first quark matter droplet in a neutron star core that consists of 
$\beta$-stable hyperonic matter being compressed during a time scale
for the spin-down or accretion.  The resulting crossover temperature 
from the quantum tunneling to the thermal activation regime is shown
to be large compared with the typical temperatures of matter in the 
core.  By comparing the obtained critical pressure with the central 
pressure of the star with maximum mass, we evaluate the range of the 
bag-model parameters, i.e., the bag constant, the QCD fine structure 
constant, and the strange quark mass, where quark matter is expected
to occur via deconfinement in the star.

In Sec.\ II, the formalism to calculate the rate of nucleation of a 
stable phase in a metastable phase via quantum tunneling is discussed 
taking into account the effects of relativity, energy dissipation, and
dynamical compressibility.  In Sec.\ III, the static properties of the
deconfinement transition are examined using a relativistic mean-field
theory of hadronic matter with and without hyperons and a bag model
for deconfined matter.  In Sec.\ IV, on the basis of the static 
properties obtained in Sec.\ III  
and the formalism described in Sec.\ II, the time needed to form a
quark matter droplet in the metastable phase of nuclear or hyperonic
matter is calculated in the overpressure regime; the critical 
overpressure and the quantum-thermal crossover temperature are also 
estimated allowing for the stellar conditions.  Conclusions are given
in Sec.\ V\@.  In the Appendix, the quark distributions inside a 
droplet are considered.

%%\newpage
%\vspace*{0.5cm}
%\begin{center}
%{\bf II. QUANTUM NUCLEATION THEORY}
%\end{center}
%\vspace*{0.5cm}
\section{QUANTUM NUCLEATION THEORY}
\label{sec:qnt}

In this section we first summarize a theory of quantum nucleation of a
stable phase in a low-temperature first-order phase transition as 
advanced by Lifshitz and Kagan [17].  We then extend this theory to a 
relativistic regime where the effective mass of a critical-size
droplet of the stable phase is comparable to the height of a potential
barrier.  Effects of energy dissipation [18] and dynamical 
compressibility [19] in the metastable phase, ignored above, are also 
built into the formalism to calculate the rate of quantum nucleation
of the new phase.  We assume in this section that the stable and 
metastable phases are electrically neutral quantum liquids with a
single component.

% h, c, k should be written.

%\vspace*{0.5cm}
%\begin{center}
%{\bf A. Lifshitz-Kagan theory} 
%\end{center}
%\vspace*{0.3cm}
\subsection{Lifshitz-Kagan theory}
\label{sec:lkt}

Quantum tunneling
nucleation of a stable phase in a first-order phase transition 
%via quantum fluctuations 
was first investigated by Lifshitz and Kagan
[17].  Their analysis gives us a basic tool for evaluating the time 
needed to form a real droplet of the stable phase at low temperatures
and at pressures in the vicinity of the phase equilibrium pressure.
At such pressures, a nucleated droplet contains a large number of 
particles so that the resulting energy gain can compensate for the
sharp energy increase in the interfacial layer.  One may thus consider
a droplet being in a virtual or real state to be a sphere
macroscopically characterized by its radius $R(t)$ and describe the 
tunneling behavior of a virtual droplet in the semiclassical 
approximation.  By assuming that the velocity of sound $c_{s}$ is 
sufficiently large compared with the velocity of the phase boundary, 
i.e., both phases are incompressible, the potential energy for a 
fluctuation of the radius $R$ may be expressed in a standard form 
%as 
\begin{equation}
   U(R)=\frac{4\pi R^{3}}{3}n_{2}(\mu_{2}-\mu_{1})
       +4\pi\sigma_{s} R^{2}\ .
\end{equation}
Here $\mu_{1} (\mu_{2})$ is the chemical potential of the metastable
(stable) phase calculated at fixed pressure $P$, $n_{2}$ is the number 
density of the stable phase, and $\sigma_{s}$ is the surface tension.
Expression (1) is derived from difference in the thermodynamic
potential at fixed chemical potential between the initial metastable 
phase and the inhomogeneous phase containing a single droplet.
As Fig.\ 1 illustrates, the potential barrier occurs between the
initial metastable state and the state with a real droplet having a 
critical radius, $R_{c}=3\sigma_{s}/n_{2}(\mu_{1}-\mu_{2})$, which 
satisfies $U(R_{c})=0$.  The height $U_{0}$ of this barrier is given
by $U_{0}=(4/27)4\pi\sigma_{s} R_{c}^{2}$.

The kinetic energy for a fluctuation of $R$ is also necessary for the 
determination of the quantum nucleation rate.  During such a 
fluctuation, the density discontinuity between the stable and
metastable phases induces a hydrodynamic mass flow in the medium 
around the droplet.  The velocity field $v(r)$ is obtained from the 
continuity equation and the boundary condition at the droplet surface
as
\begin{equation}
  v(r)=\left\{ \begin{array}{lll}
  \left(1-\displaystyle{\frac{n_{2}}{n_{1}}}\right)\dot{R}
  \left(\displaystyle{\frac{R}{r}}\right)^{2}\ ,
         & \mbox{$r\geq R\ ,$} \\
             \\
         0\ ,
         & \mbox{$r<R\ ,$}
 \end{array} \right.
\end{equation}
where ${\dot R}$ is the droplet growth rate, and $n_{1}$ is the number 
density of the metastable phase.  The kinetic energy $K(R)$ is thus 
given by
\begin{equation}
   K(R)=\frac{1}{2}M(R)\dot{R}^{2}\ , 
\end{equation}
where $M(R)$ is the effective droplet mass,
\begin{equation}
 M(R)=4\pi\rho_{1}
\left(1-\frac{n_{2}}{n_{1}}\right)^{2}R^{3}\ .
\end{equation}
Here $\rho_{1}$ is the mass density of the metastable phase.

At zero temperature, the time required to 
form a single droplet in the metastable phase may 
%thus 
be calculated within the WKB 
approximation [20] by using the Lagrangian for the fluctuating
droplet
\begin{equation}
   L(R,{\dot R})=\frac{1}{2}M(R)\dot{R}^{2}-U(R)\ .
\end{equation}
Here no energy dissipation in the medium around the droplet is
considered.  The energy $E_{0}$ for the zeroth bound state around
$R=0$ is obtained from the Bohr quantization condition
\begin{equation}
   I(E_{0})=\frac{3}{2}\pi\hbar\ ,
\end{equation}
where $I(E)$ is the action for the zero-point oscillation
\begin{equation}
   I(E)=2\int_{0}^{R_{-}}dR\sqrt{2M(R)[E-U(R)]}\ ,
\end{equation}
with the smaller classical turning radius $R_{-}$.  The corresponding
oscillation frequency $\nu_{0}$ and probability of barrier penetration
$p_{0}$ are given by
\begin{equation}
   \nu_{0}^{-1}=\left.\frac{dI}{dE}\right|_{E=E_{0}}
\end{equation}
and
\begin{equation}
   p_{0}=\exp\left[-\frac{A(E_{0})}{\hbar}\right]\ ,
\end{equation}
where $A(E)$ is the action under the potential barrier
\begin{equation}
   A(E)=2\int_{R_{-}}^{R_{+}}dR\sqrt{2M(R)[U(R)-E]}\ ,
\end{equation}
with the larger classical turning radius $R_{+}$.  The formation time
$\tau$ is finally calculated as
\begin{equation}
%   \tau_{0}=(\nu_{0}p_{0})^{-1}\ .
   \tau=(\nu_{0}p_{0})^{-1}\ .
\end{equation}

%\vspace*{0.5cm}
%\begin{center}
%{\bf B. Effect of relativity} 
%\end{center}
%\vspace*{0.3cm}
\subsection{Effect of relativity}
\label{sec:eor}

In case the potential barrier height $U_{0}$ divided by $c^{2}$ is as 
large as the effective mass $M(R_{c})$ of a droplet of critical size, 
the effect of relativity on the quantum nucleation should be taken into 
account so that $|\dot{R}|\leq c$ may be satisfied.  Since this 
situation will appear in Sec.\ IV, it is instructive to describe a 
relativistic version of the Lifshitz-Kagan theory.  Let us here assume
that the velocity field itself is small enough to ensure the 
nonrelativistic description given by Eq.\ (2).  The Lagrangian may
then be rewritten as
\begin{equation}
   L(R,{\dot R})=-M(R)c^{2}\sqrt{1-\left(\frac{\dot{R}}{c}\right)^{2}}
    +M(R)c^{2}-U(R)\ ,
\end{equation}
where $M(R)$ and $U(R)$ are given by Eqs.\ (4) and (1), respectively.

We proceed to obtain the time $\tau$ needed to form a droplet from the
Lagrangian (12) in the semiclassical approximation.  The
Hamilton-Jacobi equation is derived from the Lagrangian (12) in a 
usual way as 
\begin{equation}
  (Mc^{2})^{2}=\left(\frac{\partial S}{\partial t}+U-Mc^{2}\right)^{2}
    -\left(\frac{\partial S}{\partial R}\right)^{2}c^{2}\ ,
\end{equation}
where $S(R,t)$ is the action associated with the Lagrangian (12).  We 
perform the first quantization of Eq.\ (13) by replacing $S$ with
$\hbar/i$, and we thereby obtain a time-independent equation for the
wave function $\psi(R)$ representing the state of energy $E$ as
\begin{equation}
  \left[-\hbar^{2}c^{2}\frac{d^{2}}{dR^{2}}
  +(U-E)(2Mc^{2}+E-U)\right]\psi=0\ .
\end{equation}
Since Eq.\ (14) bears a resemblance to the nonrelativistic 
Schr{\" o}dinger equation, we can derive the semiclassical solution 
to the tunneling problem described in terms of Eq.\ (14) by following
a usual line of argument developed in the nonrelativistic case [21]. 
As a consequence, a set of equations for determining the formation
time $\tau$ is obtained in the forms analogous to Eqs.\ (6)--(11).
The energy $E_{0}$ for the zeroth bound state around $R=0$ is now 
determined from the Bohr quantization condition
\begin{equation}
   I(E_{0})=2\pi\left(m_{0}+\frac{3}{4}\right)\hbar\ .
\end{equation}
Here $I(E)$, the action for the zero-point oscillation, is rewritten
as
\begin{equation}
   I(E)=\frac{2}{c}\int_{0}^{R_{-}}dR
        \sqrt{[2M(R)c^{2}+E-U(R)][E-U(R)]}\ ,
\end{equation}
with the smaller classical turning radius $R_{-}$; $m_{0}$ is the
integer defined as 
\begin{equation}
   m_{0}=\left[\frac{I(E_{\rm min})}{2\pi\hbar}+\frac{1}{4}\right]\ ,
\end{equation}
where $E_{\rm min}$ is the maximum value of $U(R)-2M(R)c^{2}$, and 
$[\cdots]$ denotes the Gauss' notation. $E_{\rm min}$, being positive 
definite, yields the lower bound of the energy region where  
$positive$-$energy$ states occur; by these we denote the states
which ensure $2M(R)c^{2}+E-U(R)\geq0$ for arbitrary $R$.  The
zero-point oscillation frequency $\nu_{0}$ is determined by
substituting Eq.\ (16) into Eq.\ (8).  On the other hand, the 
probability $p_{0}$ of barrier penetration at $E=E_{0}$ is given by 
Eq.\ (9) in which we now use the underbarrier action
\begin{equation}
   A(E)=\frac{2}{c}\int_{R_{-}}^{R_{+}}dR
        \sqrt{[2M(R)c^{2}+E-U(R)][U(R)-E]}\ ,
\end{equation}
with the larger classical turning radius $R_{+}$.  The obtained
results for $\nu_{0}$ and $p_{0}$ lead to the formation time $\tau$
via Eq.\ (11). Equations (12)--(18) reduce to the 
nonrelativistic counterparts in the limit of $c\rightarrow\infty$.
The higher order effect of relativity acts to raise the energy level 
and hence to enhance the tunneling probability, as will be clarified 
in Sec.\ IV\@.

%\vspace*{0.5cm}
%\begin{center}
%{\bf C. Effect of energy dissipation} 
%\end{center}
%\vspace*{0.3cm}
\subsection{Effect of energy dissipation}
\label{sec:eoed}

So far it was assumed that the system adjusts adiabatically to the 
fluctuation of $R$, i.e., a virtual droplet of the stable phase 
fluctuates in a reversible way.  In a realistic situation, however, 
relaxation processes involved proceed at a finite rate; the 
density adjustment 
%redistribution of the components 
with varying $R$ is necessarily 
accompanied by appearance of excitations in the metastable phase. 
These excitations give rise to dissipation of the total energy of the 
droplet until the system reaches a complete thermodynamic equilibrium.
If the mean free path $l$ of the excitations is much smaller than
$R$, the excitations play a role in viscous transport of momentum from
the high to the low velocity region with the rate of energy
dissipation,
\begin{eqnarray}
 \frac{dE}{dt} &=&
    4\pi\eta\left(r^{2}\frac{dv^{2}}{dr}\right)_{r=R+0}
%   -24\pi\eta\int_{0}^{\infty}dr r^{2}\left(\frac{dv}{dr}\right)^{2}
  \nonumber \\ &=&
   -16\pi\eta\left(1-\frac{n_{2}}{n_{1}}\right)^{2}R{\dot R}^{2}\ ,
\end{eqnarray}
where $\eta$ is the viscosity of the metastable phase and $v(r)$ is
given by Eq.\ (2).  This energy dissipation is thus built into the 
equation of motion for the droplet derived from Eq.\ (5) or (12) as an 
Ohmic friction force
\begin{equation}
   F=-16\pi\eta\left(1-\frac{n_{2}}{n_{1}}\right)^{2}R{\dot R}\ .
\end{equation}
In the case in which $l\sim R$, however, the hydrodynamic description
of the dissipative processes ceases to be valid; considerations of the
energy dissipation require the use of the corresponding quantum
kinetic equation.  In the regime $l\gg R$ where the excitations behave
ballistically, they collide with the droplet surface and eventually 
relax in the medium far away from the droplet.  The resulting momentum
transfer dissipates the total energy of the droplet at a rate
\begin{equation}
 \frac{dE}{dt}=-16\pi\alpha\eta\left(1-\frac{n_{2}}{n_{1}}\right)^{2}
 \frac{R^{2}}{l}{\dot R}^{2}\ ,
\end{equation}
where $\alpha$ is a factor of order unity that depends on the nature
of the excitations and their interactions with the droplet surface,
and thus exerts an Ohmic friction force on the droplet:
\begin{equation}
 F=-16\pi\alpha\eta\left(1-\frac{n_{2}}{n_{1}}\right)^{2}
  \frac{R^{2}}{l}{\dot R}\ .
\end{equation}
It is to be noted that expressions (21) and (22) are based on the 
hydrodynamic equation for the velocity field, which is not applicable
to the situation considered here, and hence are not completely obvious.

The influence of the Ohmic dissipation described above on the quantum
nucleation rate was investigated by Burmistrov and Dubovskii [18]
in terms of a path-integral formalism.  This formalism was pioneered
by Caldeira and Leggett [22] to consider the dissipation effects on 
quantum tunneling in macroscopic systems; it was noted that these
effects act to reduce the tunneling probability exponentially. 
Hereafter, we 
%shall 
extend the estimation of the nucleation rate to the system of finite 
temperature $T$ with the help of path integrals, as advanced by Larkin 
and Ovchinnikov [23].  The probability $p(T)$ of formation of a droplet
in the medium undergoing the energy dissipation is then given by the 
expression similar to Eq.\ (9): 
\begin{equation}
   p=\exp\left[-\frac{A(T)}{\hbar}\right]\ .
\end{equation}
Here $A(T)$ is the extremal value of the effective action 
$S_{\rm eff}$ specified in terms of an imaginary time $\tau'$:
\begin{eqnarray}
   S_{\rm eff}[R(\tau')] &=& \int_{-\hbar\beta/2}^{\hbar\beta/2}d\tau'
   \left\{
    M(R)c^{2}\sqrt{1+\left(\frac{\dot{R}}{c}\right)^{2}}
    -M(R)c^{2}+U(R)\right.
   \nonumber \\ & & \left.
    +\frac{\eta}{4\pi}\int_{-\hbar\beta/2}^{\hbar\beta/2}d\tau''
    [\gamma(R_{\tau'})-\gamma(R_{\tau''})]^{2}\frac{(\pi/\hbar\beta)^{2}}
    {\sin^{2}\pi (\tau'-\tau'')/\hbar\beta}
    \right\}\ ,
\end{eqnarray}
where $\beta=(k_{B}T)^{-1}$ is the reciprocal temperature, $R_{\tau'}$ 
denotes $R(\tau')$, and $\gamma(R)$, the quantity determining the
nonlocal dissipation term, is evaluated from the Ohmic friction
forces (20) and (22) as 
\begin{equation}
  \gamma(R)=\left\{ \begin{array}{lll}
    \displaystyle{\frac{8\pi^{1/2}}{3}}
 \left|1-\displaystyle{\frac{n_{2}}{n_{1}}}\right|R^{3/2}\ ,
         & \mbox{$R\gg l\ ,$} \\
             \\
        2(\pi\alpha)^{1/2}
 \left|1-\displaystyle{\frac{n_{2}}{n_{1}}}\right|
     \displaystyle{\frac{R^{2}}{l^{1/2}}}\ ,
         & \mbox{$R\ll l\ .$}
 \end{array} \right.
\end{equation}
The effective action (24) is accompanied by the periodic boundary
condition $R(\hbar\beta/2)=R(-\hbar\beta/2)$.

Generally, one obtains two types of trajectories that extremize the 
effective action $S_{\rm eff}$.  One of these trajectories corresponds
to a $\tau'$-independent classical trajectory $R(\tau')=2R_{c}/3$,
along which the action amounts to $\hbar U_{0}/k_{B}T$.  The
nucleation described by this trajectory proceeds via thermal 
fluctuations.  The other trajectory depends explicitly on $\tau'$ and
denotes the nucleation occurring via quantum-mechanical fluctuations. 
This statement is evident from the fact that the corresponding
extremal action $A(T)$ reduces to $A(E_{0})$ given by Eq.\ (18) in the
limit of no energy dissipation and zero temperature. According as
$A(T)$ dominates over $\hbar U_{0}/k_{B}T$ or not, the nucleation of
the stable phase proceeds via thermal activation or via quantum 
tunneling.  There exists an abrupt transition from the quantum
tunneling to the thermal activation regime at the temperature $T_{0}$
derived from the relation
\begin{equation}
   A(T_{0})=\frac{\hbar U_{0}}{k_{B}T_{0}}\ .
\end{equation}

For the purposes of practical application, it is useful to summarize
the results for $A(T)$ in the ballistic $(R\ll l)$ and nonrelativistic 
regime as given in Ref.\ [18].  In the case of weak dissipation, the 
dissipative processes add a relatively small quantity $A_{1}(T)$ to 
$A(E_{0})$; the analytic expressions for $A(E_{0})$ and $A_{1}(T)$ are 
given at $E_{0}=0$ by
\begin{equation}
A(E_{0}=0)=\frac{5\sqrt{2}\pi^{2}}{16}\left|1-\frac{n_{2}}{n_{1}}
         \right|\sqrt{\sigma_{s}\rho_{1}}R_{c}^{7/2}
\end{equation}
and
\begin{equation}
A_{1}(T)=4\pi\alpha\left(1-\frac{n_{2}}{n_{1}}\right)^{2}
    \frac{\eta R_{c}^{4}}{l}\left[\frac{J_{2}}{\pi}+O(T^{2})\right]\ ,
\end{equation}
with $J_{2}\approx 1.49$.  For temperatures up to the crossover point 
$T_{0}$, not only does $A(E_{0}=0)$ retain its form (27) but also the 
term of order $T^{2}$ in Eq.\ (28) is negligible.  For normal Fermi 
excitations in the medium, we obtain $\eta/l\sim n_{1}p_{F}$ with the 
Fermi momentum $p_{F}=\hbar(3\pi^{2}n_{1})^{1/3}$.  Here it is 
instructive to note that in the case of weak dissipation, the Fermi 
velocity $v_{F}=n_{1}p_{F}/\rho_{1}$ is negligibly small compared with 
the characteristic velocity $\sqrt{2U_{0}/M(R_{c})}$ of a virtual 
droplet moving under the potential barrier.  In the opposite case of 
strong dissipation [$v_{F}\gg\sqrt{2U_{0}/M(R_{c})}$ for normal Fermi 
excitations], the dissipation effects control the dynamics of the 
droplet in configuration space and thus allow one to leave out the 
kinetic term in the effective action (24).  The corresponding 
expression for $A(T)$ reads
\begin{equation}
A(T)=4\pi\alpha\left(1-\frac{n_{2}}{n_{1}}\right)^{2}
    \frac{\eta R_{c}^{4}}{l}s(T)\ ,
\end{equation}
where $s(T)$, the normalized underbarrier action calculated along the
extremal trajectory, is $\approx 1.3$ in the temperature range 
including $T\leq T_{0}$.

%\vspace*{0.5cm}
%\begin{center}
%{\bf D. Effect of dynamical compressibility} 
%\end{center}
%\vspace*{0.3cm}
\subsection{Effect of dynamical compressibility}
\label{sec:eodc}

In a system where the metastable phase is more or less compressible,
a finite compressibility acts to reduce a portion of the liquid
taking part in the hydrodynamic mass flow and hence the overall
kinetic energy, leading to an exponential increase in the quantum 
nucleation rate.  These effects were considered by Korshunov [19]
in the nonrelativistic regime ($c\gg|\dot R|$) by taking account of
the time dependence of the velocity field.  We formally extend his 
theory to the relativistic regime by replacing the kinetic term in the
effective action (24) with
\begin{eqnarray}
   S_{M}[R(\tau')] &=& \int_{-\hbar\beta/2}^{\hbar\beta/2}d\tau'
   \left[-1+\sqrt{1+\left(\frac{\dot{R}}{c}\right)^{2}}\right]
   \left(\frac{c}{{\dot R}}\right)^{2}
       \nonumber \\ & &
   \times\int_{R(\tau')}^{\infty}dr 4\pi\rho_{1}r^{2}
   \left[\left(\frac{\partial\varphi}{\partial r}\right)^{2}
         +\frac{1}{c_{s}^{2}}
   \left(\frac{\partial\varphi}{\partial\tau'}\right)^{2}\right]\ ,
\end{eqnarray}
where $\varphi(r,\tau')$ is the velocity potential assumed to be
spherically symmetric.  The form of $S_{M}$ adopted here
is available when the flow velocity $|v|$ is much smaller than
$c_{s}$.  The extremalization of $S_{M}$ in $\varphi$ results in the
wave equation
\begin{equation}
  \frac{\partial^{2}\varphi}{\partial\tau'^{2}}+
  \frac{c_{s}^{2}}{r^{2}}\frac{\partial}{\partial r}
  \left(r^{2}\frac{\partial\varphi}{\partial r}\right)=0\ .
\end{equation}
Equation (31) can be solved under the boundary condition at the
droplet surface based on the assumption that the stable phase is
incompressible:
\begin{equation}
\left(\frac{n_{2}}{n_{1}}-1\right){\dot R}
  =-\left.\frac{\partial\varphi(r,\tau')}{\partial r}
    \right|_{r=R(\tau')}\ .
\end{equation}
It is obvious from Eq.\ (30) that $S_{M}$ reduces to the kinetic term
in Eq.\ (24) (hereafter defined as $S_{K}$) in the incompressible
limit $(|{\dot R}|\ll c_{s})$.  The difference $S_{C}=S_{M}-S_{K}$ has
a nonlocal character similar to the dissipation term in
Eq.\ (24), as shown in Ref.\ [19].

The compressibility term $S_{C}$, if the trajectory involved is taken 
to be the extremal one obtained at $T=0$ and $E_{0}=0$ in the 
nonrelativistic and dissipationless limit, leads to the expression
[19]
\begin{equation}
A_{C}(T)=-490\left(\frac{3\pi}{4}\right)^{6}
 \left(\frac{R_{c}}{3}\right)^{9}
 \left(1-\frac{n_{2}}{n_{1}}\right)^{4}
 \frac{\rho_{1}^{2}}{\sigma_{s} c_{s}}	
 \left(\frac{k_{B}T}{\hbar}\right)^{4}\ .
\end{equation}
Expression (33) describes just the principal term in the expansion
with respect to $T/T_{0}$ and $\sqrt{2U_{0}/M(R_{c})}/c_{s}$.  It is
noteworthy that $A_{C}(T)$ is proportional to $(T/T_{0})^{4}$ for
$T\ll T_{0}$.  This behavior, attributable to the emission of the
sound wave in the course of a fluctuation of the droplet radius, 
persists in the case in which the condition 
$\sqrt{2U_{0}/M(R_{c})}\ll c_{s}$ is violated, the energy dissipation
takes effect, and/or the droplet surface moves relativistically, as 
long as the time of flight of the extremal trajectory is finite for 
$T=0$.  Here we refrain from proving such finiteness explicitly, but
this is a feature generally held by problems with a mass varying with 
$R$.

%%\newpage
%\vspace*{0.5cm}
%\begin{center}
%{\bf 3. STATIC DECONFINEMENT TRANSITION}
%\end{center}
%\vspace*{0.5cm}
\section{STATIC DECONFINEMENT TRANSITION}
\label{sec:sdt}

We now estimate at what pressure hadronic matter in $\beta$
equilibrium undergoes a static deconfinement transition at zero 
temperature subject to flavor conservation.\footnote{Hereafter, we
take units in which $\hbar=c=k_{B}=1$.} For this purpose, we first 
describe the bulk properties of the $\beta$-stable, zero-temperature
hadronic matter with and without hyperons within the framework of a 
relativistic nuclear field theory [24].  Its Lagrangian density,
denoting the interactions between baryons by the exchange of 
${\cal M}$ mesons (${\cal M}=\sigma,\omega,\rho$), is given by
\begin{eqnarray}
{\cal L} &=& \sum_{B}{\bar\psi}_{B}\left(i\gamma_{\mu}\partial^{\mu}-
m_{B}+g_{\sigma B}\sigma-g_{\omega B}\gamma_{\mu}\omega^{\mu}
-\frac{1}{2}g_{\rho B}\gamma_{\mu}
\mbox{\boldmath $\tau\cdot\rho$}^{\mu}\right)\psi_{B}
   \nonumber \\ & &
+\frac{1}{2}(\partial_{\mu}\sigma\partial^{\mu}\sigma-
m_{\em\sigma}^{2}\sigma^{2})-\frac{1}{4}\omega_{\mu\nu}\omega^{\mu\nu}
+\frac{1}{2}m_{\omega}^{2}\omega_{\mu}\omega^{\mu}
-\frac{1}{4}\mbox{\boldmath $\rho$}_{\mu\nu}\mbox{\boldmath 
$\cdot\rho$}^{\mu\nu}
+\frac{1}{2}m_{\rho}^{2}\mbox{\boldmath $\rho$}_{\mu}
\mbox{\boldmath $\cdot\rho$}^{\mu}
   \nonumber \\ & &
-{\cal U(\sl\sigma)}+\sum_{l}{\bar\psi}_{l}
(i\gamma_{\mu}\partial^{\mu}-m_{l})\psi_{l}\ ,
\end{eqnarray}
with 
\begin{eqnarray}
\omega_{\mu\nu} &=& \partial_{\mu}\omega_{\nu}
               -\partial_{\nu}\omega_{\mu}\ , \nonumber \\
\mbox{\boldmath $\rho$}_{\mu\nu} &=& \partial_{\mu}
      \mbox{\boldmath $\rho$}_{\nu}
     -\partial_{\nu}\mbox{\boldmath $\rho$}_{\mu}\ .
  \nonumber
\end{eqnarray}
Here $\psi_{B(l)}$ is the Dirac spinor for baryons 
$B=p,n,\Lambda,\Sigma^{+},\Sigma^{-},\Sigma^{0},\Xi^{-},\Xi^{0}$ 
(for leptons $l=e^{-},\mu^{-}$), $\mbox{\boldmath $\tau$}$ is the
isospin operator, $g_{{\cal M}B}$ is the coupling constant between
$B$ baryons and ${\cal M}$ mesons, and $m_{i}$ is the rest mass of 
the particle of species $i$ ($i=B,{\cal M},l$).  The potential 
$\cal U(\sigma)$, denoting the self-interactions of the scalar field
and playing a role in reproducing the empirical nuclear 
incompressibility, takes a specific form,
\begin{equation}
{\cal U(\sl\sigma)}=\frac{1}{3}a_{1}m_{n}(g_{\sigma N}\sigma)^{3}
                +\frac{1}{4}a_{2}(g_{\sigma N}\sigma)^{4}\ ,
\end{equation}
where $N$ represents the nucleons ($N=n,p$).

From the Lagrangian density (34) we obtain a couple of the 
Euler-Lagrange equations with respect to the meson and baryon fields.
In the mean field approximation adopted here, the meson fields are 
replaced by their expectation values and are determined from the 
classical field equations [14]
\begin{equation}
m_{\omega}^{2}\omega_{0}=\sum_{B}g_{\omega B}n_{B}\ ,
\end{equation}
\begin{equation}
m_{\rho}^{2}\rho_{30}=\sum_{B}g_{\rho B}\tau_{3B}n_{B}\ ,
\end{equation}
\begin{equation}
m_{\sigma}^{2}\sigma=-\frac{d{\cal U}(\sigma)}{d\sigma}
+\frac{1}{2\pi^{2}}\sum_{B}g_{\sigma B}M_{B}^{3}
  \left[t_{B}\sqrt{1+t_{B}^{2}}
  -\ln\left(t_{B}+\sqrt{1+t_{B}^{2}}\right)\right]\ ,
\end{equation}
where $n_{B}$ is the number density of $B$ baryons, 
$M_{B}=m_{B}-g_{\sigma B}\sigma$ is the effective mass of the baryon, 
$\tau_{3B}$ is the third component of isospin of the baryon, and 
$t_{B}=k_{F,B}/M_{B}$ with the corresponding Fermi wave number 
$k_{F,B}=(3\pi^{2}n_{B})^{1/3}$.  Here the meson fields are constant
for any space and time since the considered system is static and
uniform.  This system is also isotropic, leading to
$\mbox{\boldmath $\omega$}=\mbox{\boldmath $\rho_{3}$}=0$.
We have assumed, furthermore, that the charged components of $\rho$
mesons as well as the pions and the kaons have vanishing expectation 
values, as is normally the case.

The total energy density $\varepsilon_{\rm tot}$ and pressure $P$ for
the hadronic system are then calculated from the energy-momentum
tensor derived from the Lagrangian density (34) as a sum of the
hadronic term and the leptonic term [14].  The resulting expression
for $\varepsilon_{\rm tot}$ is
\begin{equation}
\varepsilon_{\rm tot}=\varepsilon_{H}+\varepsilon_{L}\ , 
\end{equation}
with
\begin{eqnarray}
\varepsilon_{H} &=& {\cal U(\sigma)}+\frac{1}{2}m_{\sigma}^{2}
  \sigma^{2}+\frac{1}{2}m_{\omega}^{2}\omega_{0}^{2}+
  \frac{1}{2}m_{\rho}^{2}\rho_{30}^{2}
   \nonumber \\ & &
  +\frac{1}{8\pi^{2}}\sum_{B}M_{B}^{4}
  \left[(2t_{B}^{2}+1)t_{B}\sqrt{1+t_{B}^{2}}
      -\ln\left(t_{B}+\sqrt{1+t_{B}^{2}}\right)\right]\ ,   
\end{eqnarray}
\begin{equation}
\varepsilon_{L}=\frac{1}{8\pi^{2}}\sum_{l}m_{l}^{4}
  \left[(2t_{l}^{2}+1)t_{l}\sqrt{1+t_{l}^{2}}
      -\ln\left(t_{l}+\sqrt{1+t_{l}^{2}}\right)\right]\ ,   
\end{equation}
where $t_{l}=k_{F,l}/m_{l}$ with the Fermi wave number of $l$ leptons
$k_{F,l}=(3\pi^{2}n_{l})^{1/3}$.
%\begin{eqnarray}
% P_{\rm had} &=& -{\cal U(\sigma)}-\frac{1}{2}m_{\sigma}^{2}
%  \sigma^{2}+\frac{1}{2}m_{\omega}^{2}\omega_{0}^{2}+
%  \frac{1}{2}m_{\rho}^{2}\rho_{0}^{2}
%   \nonumber \\ & &
%  -\frac{1}{8\pi^{2}}\sum_{B}M_{B}^{4}
%  \left[(2t_{B}^{2}+1)t_{B}\sqrt{1+t_{B}^{2}}
%      -\ln(t_{B}+\sqrt{1+t_{B}^{2}})\right]   
%   \nonumber \\ & &
%  -\frac{1}{8\pi^{2}}\sum_{l}m_{l}^{4}
%  \left[(2t_{l}^{2}+1)t_{l}\sqrt{1+t_{l}^{2}}
%      -\ln(t_{l}+\sqrt{1+t_{l}^{2}})\right]   
%   \nonumber \\ & &
%  +\sum_{B}n_{B}\mu_{B}-\sum_{l}n_{l}\mu_{l}\ ,
%\end{eqnarray}
The pressure is obtained as
\begin{equation}
  P=P_{H}+P_{L}\ ,
\end{equation}
with
\begin{equation}
P_{H}=-\varepsilon_{H}
%  +m_{\omega}^{2}\omega_{0}^{2}+m_{\rho}^{2}\rho_{30}^{2}
  +\sum_{B}n_{B}\mu_{B}\ ,
\end{equation}
\begin{equation}
P_{L}=-\varepsilon_{L}+\sum_{l}n_{l}\mu_{l}\ , 
\end{equation}
where 
\begin{equation}
\mu_{B}=\sqrt{k_{F,B}^{2}+M_{B}^{2}}+g_{\omega B}\omega_{0}
        +g_{\rho B}\tau_{3B}\rho_{30}
\end{equation}
is the chemical potential of $B$ baryons, and  
\begin{equation}
\mu_{l}=\sqrt{k_{F,l}^{2}+m_{l}^{2}}
\end{equation}
is the chemical potential of $l$ leptons.  Here electrons and muons
are looked upon as ideal gases; the density region considered here is 
high enough for the kinetic energy to dominate over the energy induced
by Coulomb interactions.
%We in turn obtain the baryon chemical potential as
%\begin{equation}
%\mu_{\rm had}=\frac{\varepsilon_{\rm tot}+P}{n_{b}}\ ,
%\end{equation}
%where $n_{b}=\sum_{B}n_{B}$ 
%is the total baryon density of hadronic matter.
The pressure $P$ is in turn related to the total energy density 
$\varepsilon_{\rm tot}$, equivalent to the mass density $\rho$, via
\begin{equation}
 P=n_{b}^{2}\frac{\partial}{\partial n_{b}}
 \left(\frac{\varepsilon_{\rm tot}}{n_{b}}\right)\ ,
\end{equation}
where $n_{b}=\sum_{B}n_{B}$ is the total baryon density of hadronic 
matter.

The parameters $g_{\sigma B}$, $g_{\rho B}$, $g_{\omega B}$, $a_{1}$, 
and $a_{2}$ contained in Eqs.\ (34) and (35) have been taken from the
values determined by Glendenning and Moszkowski [16].  The properties
of saturated nuclear matter adopted in Ref.\ [16] are characterized by
the binding energy per nucleon (16.3 MeV), the saturation density 
(0.153 fm$^{-3}$), the symmetry energy (32.5 MeV), the
incompressibility (300 MeV), and the effective nucleon mass 
($0.7m_{n}$).  In such a way as to reproduce these properties, the 
nucleon-meson coupling constants and the parameters determining the 
strength of the $\sigma$ self-interactions were chosen as
$g_{\sigma N}/m_{\sigma}=3.434$ fm, $g_{\omega N}/m_{\omega}=2.674$ fm,
$g_{\rho N}/m_{\rho}=2.100$ fm, $a_{1}=0.00295$, and $a_{2}=-0.00107$.
The hyperon-meson couplings, denoted by the ratios
\begin{equation}
x_{\sigma H}=\frac{g_{\sigma H}}{g_{\sigma N}}\ ,\  
x_{\omega H}=\frac{g_{\omega H}}{g_{\omega N}}\ ,\  
x_{\rho H}=\frac{g_{\rho H}}{g_{\rho N}}\ ,
\end{equation}
were determined mainly from the empirical $\Lambda$ binding energy
(28 MeV) in saturated nuclear matter as $x_{\sigma H}=x_{\rho H}=0.6$
and $x_{\omega H}=0.653$.  Here the coupling constant ratios are
assumed to be the same for all hyperon species.  It was noted that the
ratios thus chosen are compatible with the observational lower bound
of the maximum neutron star mass and with the upper bound of 
$x_{\sigma H}$ stemming from the fit to hypernuclear levels.

In the determination of the equilibrium composition of the hadronic
system at a given $n_{b}$, we need not only Eqs.\ (36)--(38), but also
additional constraints.  These are the charge neutrality condition
\begin{equation}
   \sum_{B}q_{B}n_{B}-n_{e}-n_{\mu}=0\ ,
\end{equation}
where $q_{B}$ is the electric charge of the baryon of species $B$, and 
the $\beta$-equilibrium conditions
\begin{eqnarray}
 & & \mu_{B}=\mu_{n}-q_{B}\mu_{e}\ , \\ & &
   \mu_{\mu}=\mu_{e}\ .
\end{eqnarray}
We have thus
%thereby 
calculated the fraction $Y_{i}=n_{i}/n_{b}$ of $i$ 
particles ($i=B,l$) in matter with and without hyperons; the results 
have been plotted as a function of $n_{b}$ in Fig.\ 2.  We can observe 
in this figure that the lepton fractions in hyperonic matter are 
drastically reduced by the appearance of $\Sigma^{-}$ hyperons in the
density region $n_{b}\gtrsim 2n_{0}$.  In Fig.\ 3 we have depicted the 
equations of state of hadronic matter as calculated allowing for and 
ignoring the presence of hyperons.  For comparison we have also
included the equations of state of equilibrium nuclear matter in the 
many-body calculations [25,26] based on realistic nucleon-nucleon 
interactions.  We thus see that the presence of hyperons at high 
densities as predicted from the relativistic mean-field theory
softens the equation of state considerably.  The
relativistic field theory used here automatically ensures the
causality condition that the sound velocity
\begin{equation}
   c_{s}=\left(\frac{\partial P}{\partial\rho}
        \right)^{1/2}=\left(\frac{n_{b}}{\rho+P}
        \frac{\partial P}{\partial n_{b}}\right)^{1/2}
\end{equation}
be confined within the velocity of light.
% leading to a reduction in the maximum mass as will be stated below.

We turn to the description of the bulk properties of uniform quark 
matter, deconfined from the $\beta$-stable hadronic matter mentioned 
above, by using an MIT bag model.  We begin with the thermodynamic 
potential of $q$ quarks, where $q=u,d$, and $s$ denote up, down, and 
strange quarks, expressed as a sum of the kinetic term and the 
one-gluon-exchange term [2,27,28]
\begin{eqnarray}
\Omega_{q} &=& -\frac{1}{4\pi^{2}}
\left[\mu_{q}(\mu_{q}^{2}-m_{q}^{2})^{1/2}
\left(\mu_{q}^{2}-\frac{5}{2}m_{q}^{2}\right)
+\frac{3}{2}m_{q}^{4}
\ln\frac{\mu_{q}+(\mu_{q}^{2}-m_{q}^{2})^{1/2}}{m_{q}}\right]
   \nonumber \\ & &
+\frac{\alpha_{s}}{2\pi^{3}}\left\{3\left[\mu_{q}
(\mu_{q}^{2}-m_{q}^{2})^{1/2}-m_{q}^{2}\ln\frac{\mu_{q}+
(\mu_{q}^{2}-m_{q}^{2})^{1/2}}{m_{q}}\right]^{2}\right.
   \nonumber \\ & & \left.\frac{}{}
-2(\mu_{q}^{2}-m_{q}^{2})^{2}\right\}\ ,
\end{eqnarray}
where $m_{q}$ and $\mu_{q}$ are the $q$ quark rest mass and chemical 
potential, respectively, and $\alpha_{s}$ denotes the QCD fine
structure constant.  For $u$ and $d$ quarks, the ratios of the mass to
the chemical potential are negligibly small; we thus use the following
expressions in the massless limit:
\begin{equation}
\Omega_{u}=-\frac{\mu_{u}^{4}}{4\pi^{2}}
\left(1-\frac{2\alpha_{s}}{\pi}\right)\ ,\  
\Omega_{d}=-\frac{\mu_{d}^{4}}{4\pi^{2}}
\left(1-\frac{2\alpha_{s}}{\pi}\right)\ .
\end{equation}
The number density $n_{q}$ of $q$ quarks is in turn related to 
$\Omega_{q}$ via
\begin{equation}
 n_{q}=-\frac{\partial\Omega_{q}}{\partial \mu_{q}}\ .
\end{equation}

Then, the total energy density for the quark system is
\begin{equation}
\varepsilon_{\rm tot}=\varepsilon_{Q}+\varepsilon_{L}\ , 
\end{equation}
with
\begin{eqnarray}
\varepsilon_{Q} &=& \sum_{q}(\Omega_{q}+\mu_{q}n_{q})+b
   \nonumber \\ &=&
  \left(1-\frac{2\alpha_{s}}{\pi}\right)
  \frac{3}{4\pi^{2}}(\mu_{u}^{4}+\mu_{d}^{4})
  +\frac{3m_{s}^{4}}{8\pi^{2}}[x_{s}\eta_{s}(2x_{s}^{2}+1)
                    -\ln(x_{s}+\eta_{s})]
   \nonumber \\ & &
  -\frac{\alpha_{s}m_{s}^{4}}{2\pi^{3}}
  \{2x_{s}^{2}(x_{s}^{2}+2\eta_{s}^{2})
  -3[x_{s}\eta_{s}+\ln(x_{s}+\eta_{s})]^{2}\}+b\ ,
\end{eqnarray}
where $x_{s}=\sqrt{\mu_{s}^{2}-m_{s}^{2}}/m_{s}$, 
$\eta_{s}=\sqrt{1+x_{s}^{2}}$, $\varepsilon_{L}$ is given by 
Eq.\ (41) and $b$ is the energy density difference between the 
perturbative vacuum and the true vacuum, i.e., the bag constant.  The
total baryon density is now given by 
\begin{equation}
n_{b}=\frac{1}{3}\sum_{q}n_{q}\ .
\end{equation}
By using the thermodynamic relation, we obtain the pressure
\begin{equation}
P=P_{Q}+P_{L}\ , 
\end{equation}
with
\begin{equation}
P_{Q}=-\sum_{q}\Omega_{q}-b\ ,
\end{equation}
where $P_{L}$ is given by Eq.\ (44).

The parameters $m_{s}$, $\alpha_{s}$, and $b$ are basically obtained 
from the fits to light-hadron spectra.  However, the values of these 
parameters yielded by such fits do not correspond directly to the
values appropriate to bulk quark matter [28].  Taking account of 
possible uncertainties, we set 
$50$ MeV fm$^{-3}\leq b\leq$ 200 MeV fm$^{-3}$,
$0\leq m_{s}\leq$ 300 MeV, and $0\leq\alpha_{s}\leq 1$. 
The ranges of $\alpha_{s}$ and $m_{s}$ so chosen are consistent with
%the empirical values inferred 
the results of Barnett $et$ $al$. [29] that are inferred from
experiments and renormalized at an energy scale $\sim1$ GeV of
interest here.

In obtaining the composition of the quark system, we take note of its
relation via flavor conservation with that of the $\beta$-stable
hadronic system [30]:
\begin{equation}
\left(\begin{array}{c} Y_{u} \\ Y_{d} \\ Y_{s} \end{array}\right)
=\left(\begin{array}{cccccccc} 
2 & 1 & 1 & 2 & 1 & 0 & 1 & 0 \\ 
1 & 2 & 1 & 0 & 1 & 2 & 0 & 1 \\ 
0 & 0 & 1 & 1 & 1 & 1 & 2 & 2 \end{array}\right)
\left(\begin{array}{c} 
Y_{p} \\ Y_{n} \\ Y_{\Lambda} \\ Y_{\Sigma^{+}} \\ 
Y_{\Sigma^{0}} \\ Y_{\Sigma^{-}} \\ Y_{\Xi^{0}} \\ Y_{\Xi^{-}}
\end{array}\right)\ ,
\end{equation}
where $Y_{q}=n_{q}/n_{b}$ with $n_{b}$ given by Eq.\ (58) is the 
fraction of $q$ quarks in the deconfined phase.  The absence of 
leptonic weak processes and the charge neutrality condition ensure 
that the fraction $Y_{l}$ of $l$ leptons remains unchanged before and 
after deconfinement.

Let us now proceed to obtain the pressure $P_{0}$ of the static 
deconfinement transition from the relation
\begin{equation}
 \mu_{b}(P_{0})|_{\rm hadronic~ phase}
=\mu_{b}(P_{0})|_{\rm deconfined~ phase}\ ,
\end{equation}
where
\begin{equation}
 \mu_{b}=\frac{\varepsilon_{\rm tot}+P}{n_{b}}\ ,
\end{equation}
with the corresponding total energy density $\varepsilon_{\rm tot}$
and total baryon density $n_{b}$ is the baryon chemical potential at
fixed $P$.  We have 
%therefrom 
calculated the pressures $P_{0}$ for the parametric combinations 
between $50$ MeV fm$^{-3}\leq b\leq$ 200 MeV fm$^{-3}$,
$m_{s}=0, 150, 300$ MeV, and $0\leq\alpha_{s}\leq 1$, allowing for and
ignoring the presence of hyperons in the hadronic phase.  Figure 4 
illustrates the resultant contour plots of the pressures $P_{0}$ on
the $b$ versus $\alpha_{s}$ plane.  As specific values of $P_{0}$
shown in Fig.\ 4, we have chosen the pressure $P_{\Sigma^{-}}$ at
which $\Sigma^{-}$ hyperons appear in the hadronic phase, the central 
pressure $P_{\rm max}$ of the maximum-mass ($M_{\rm max}$) neutron star 
having a hadronic matter core, and the central pressure $P_{1.4}$ of
the star (not including quark matter) with canonical mass 
1.4$M_{\odot}$; these values are tabulated in Table I\@.
$P_{\Sigma^{-}}$ can be determined from the relation coming from 
Eq.\ (50):
\begin{equation}
  \mu_{n}=\mu_{\Sigma^{-}}|_{n_{\Sigma^{-}}=0}-\mu_{e}\ ,
\end{equation}
where $\mu_{\Sigma^{-}}|_{n_{\Sigma^{-}}=0}$ denotes the 
energy of the lowest state for a
%effective mass of a single 
$\Sigma^{-}$ hyperon. 
%in nuclear matter.  
The 
tabulated values of $P_{\rm max}$ and $P_{1.4}$ have been calculated 
from the structure of the star whose core consists of $\beta$-stable 
nuclear matter or hyperonic matter.  In these calculations, the 
gravitational mass of the star observed by a distant spectator has
been obtained as a function of the central pressure or mass density of
the star from the general-relativistic equation of hydrostatic balance
in the nonrotating configuration, i.e., the Tolman-Oppenheimer-Volkoff 
equation [31], and from the corresponding equation of state. 
The equation of state used here has been extrapolated to lower 
densities appropriate to matter in the crust made up of a lattice of 
nuclei embedded in a sea of electrons and, when present, in a sea of 
neutrons.  Such extrapolations make only a negligible difference in 
determining the stellar structure, since the crust contains mass of 
$\sim 0.01 M_{\odot}$, confined within a few percent of the total mass
[32].

The effect of a finite $s$ quark mass acts to destabilize the quark
system and hence to lower the contours of various $P_{0}$
on the $b$ versus $\alpha_{s}$ plane, as shown in Figs.\ 4(b)--4(d).
We have found that for $m_{s}\lesssim200$ MeV, such lowering is fairly
small compared with the adopted range of $b$ and $\alpha_{s}$.  Thus,
we shall generally take $m_{s}$ to be zero in describing the bulk
properties of quark matter.  We also observe in Figs.\ 4(a) and 4(b)
that the contours of $P_{0}=P_{\rm max}$ and $P_{0}=P_{1.4}$ reveal
only a weak dependence on the presence of hyperons.  This result
implies that within the confines of the present models for
matter, the degree of strangeness-induced stabilization
of the hadronic phase compensates for that of the deconfined phase
as long as $m_{s}\lesssim200$ MeV\@.

For the parameters $\alpha_{s}=0.4$, $b=100$ MeV fm$^{-3}$, and 
$m_{s}=0$, we have depicted in Fig.\ 5 the baryon chemical potentials
and total energy densities for three electrically neutral bulk phases.
The first phase consists of
$\beta$-stable hadronic matter 
with or without hyperons. The second phase is composed of 
the matter deconfined therefrom according to Eq.\ (61). 
Lastly, for comparison, we consider 
a phase of $u,d$, and $s$ quark matter in $\beta$ equilibrium: 
$\mu_{d}=\mu_{u}+\mu_{e}$ and $\mu_{d}=\mu_{s}$.  Hereafter, we shall 
refer to this $\beta$-stable quark matter as strange matter.\footnote{
%Suppose that a droplet of the deconfined matter forms and grows into
%bulk matter in a neutron star. Then, it would be converted into
%strange matter via weak processes. The resulting state of the star, if
%the strange matter being surrounded by the hadronic phase, would
%remain stable or partly change into the quark-hadron mixed phase.
Although strange matter is not related to the deconfinement transition
of interest here, it would occur once a droplet of the deconfined 
matter forms and grows into bulk matter in a neutron star.  This is 
because the weak process $u+d\rightarrow u+s$ converts the bulk
deconfined matter into strange matter during the growth of the droplet 
[1].} 
It is 
instructive to note that strange matter composed of massless quarks 
satisfies $n_{u}=n_{d}=n_{s}$ and $n_{e}=n_{\mu}=0$.  The upper 
crossing shown in Fig.\ 5(a) [5(b)] denotes the deconfinement 
pressure $P_{0}$, at which the total baryon density changes as can be 
seen from the Maxwell equal-area construction plotted in Fig.\ 5(c) 
[5(d)].  The density discontinuity for hyperonic matter is 
significantly smaller than that for nuclear matter, a feature with 
immediate relevance to the nucleation of the deconfined phase as will 
be discussed in Sec.\ IV\@.  We likewise observe that the presence of 
hyperons in the hadronic phase suppresses the chemical potential
difference between deconfined and strange matter.  This is because the 
dominance of $\Lambda$ hyperons in hyperonic matter renders the quark 
fraction $Y_{q}$ in deconfined matter similar to that in strange
matter.

We conclude this section by indicating that thermal effects on the 
static properties of the deconfinement transition may be safely
omitted at low temperatures appropriate to neutron star cores.  The 
zero-temperature models for hadronic and quark matter, described
above, hold over an even wider
%a fairly wide 
range of temperature, $T<10$ MeV\@.  This is 
because the relative chemical potentials $\mu_{i}-\mu_{i}|_{n_{i}=0}$
%(not including the rest masses) 
of $i$ particles ($i=l,B,q$), except for minor components with
particle fraction $Y_{i}\lesssim 0.1$ ($i=B$) or 
$\lesssim 0.001$ ($i=l,q$), take on a value of 
$\sim30$--600 MeV, sufficiently large compared with the temperature.
This temperature range will thus be considered below.

%%\newpage
%\vspace*{0.5cm}
%\begin{center}
%{\bf IV. DYNAMICAL DECONFINEMENT TRANSITION}
%\end{center}
%\vspace*{0.5cm}
\section{DYNAMICAL DECONFINEMENT TRANSITION}
\label{sec:ddt}

In this section we first construct the effective action for the 
fluctuational development of a quark matter droplet in the metastable 
phase of hadronic matter with and without hyperons.  We thereby
estimate the time necessary for the quantum-mechanical formation of
the first droplet in a neutron star as a function of pressure.
The critical overpressure is determined in such a way as to make
this formation time comparable to the time scale for the compression
due to the star's spin-down or accretion; at this overpressure, the 
quantum-thermal crossover temperature is estimated.  We finally
determine the window of the bag-model parameters $b$, $\alpha_{s}$, 
and $m_{s}$ where quark matter is expected to nucleate in the star.

%\vspace*{0.5cm}
%\begin{center}
%{\bf A. Effective action}
%\end{center}
%\vspace*{0.3cm}
\subsection{Effective action}
\label{sec:ea}

The quantum nucleation theory described in Sec.\ II is directly 
applicable to the case in which the metastable and stable phases are
%consisting of
electrically neutral quantum liquids with a single component. The 
deconfinement phase transition of interest here, however, involves 
matter with multiple components such as baryons, leptons, and quarks. 
Moreover, the inhomogeneous state of matter including a single quark
matter droplet leads inevitably to a violation of the local charge 
neutrality, a property that is inherent in the systems (e.g., nuclei)
where strong interactions compete with Coulomb interactions.  It is 
thus necessary to generalize the quantum nucleation theory by taking 
into account the multiplicity of components and the electrostatic 
energy.

%\vspace*{0.5cm}
%\begin{center}
%1. Nuclear matter
%\end{center}
%\vspace*{0.3cm}
\subsubsection{Nuclear matter}
\label{sec:nm}

Let us now construct the effective action, leading to the rate of 
quantum nucleation of quark matter in the case in which the metastable
phase consists of $\beta$-stable nuclear matter, by following a line
of argument of Refs.\ [11,12].  We begin with the potential energy 
$U(R)$ which is defined in the limit of $c_{s}\gg |{\dot R}|$ as the 
minimum work needed to form a quark matter droplet of radius $R$ in
the metastable phase.  Pressure equilibrium between quarks and
nucleons embedded in a roughly uniform sea of leptons induces the 
excess positive charge inside the droplet, because the baryon density
of the quark phase is larger than that of the hadronic phase by 
$\sim$0.1--0.7 fm$^{-3}$.  The charged components are in turn
distributed in such a way as to screen the droplet charge.  The 
screening efficiency of $i$ particles ($i=e, \mu, p, u, d$) may be 
partially measured from their Thomas-Fermi screening length
\begin{equation}
  \lambda_{{\rm TF},i}=\left(\frac{1}{4\pi q_{i}^{2}e^{2}}
      \frac{\partial \mu_{i}}{\partial n_{i}}\right)^{1/2}\ ,
%  _{n_{j}, j\neq i}\ ,
\end{equation}
where $q_{i}$ is the electric charge of the particle of species $i$.
For the droplet sizes of interest, $R\lesssim 10$ fm, we can assume the 
fluid of protons to be homogeneous since 
$\lambda_{{\rm TF},p}\gtrsim 10$ fm in the considered density range
$n_{b}\gtrsim 4n_{0}$.  The other components, whose screening lengths 
are confined within 10 fm, deviate more or less from uniformity.  On
the other hand, the flavor conservation, 
%retaining 
holding fixed the ratio of the 
number of $u$ quarks to that of $d$ quarks inside the droplet, makes
the screening action by quarks ineffective, since $u$ and $d$ quarks 
have electric charge with opposite sign.  We assume that $u$ and $d$ 
quarks are distributed uniformly; it turns out that the quark
screening, i.e., the static compressibility of the fluid of quarks,
does not contribute significantly to a reduction of the potential
energy of the droplet (see the Appendix).

For the initial metastable phase of nuclear matter in $\beta$
equilibrium under pressure $P$, we can obtain the total energy density 
$\varepsilon_{{\rm tot},H}$, the baryon density $n_{b,H}$, and the 
baryon chemical potential $\mu_{b,H}$ by using Eqs.\ (39), (42), and 
(63).  We then express the energy density for the inhomogeneous phase 
containing a single droplet as the sum of bulk, interfacial, and
Coulomb terms:
\begin{eqnarray}
\varepsilon_{D}(r)&=&\theta(R-r)\varepsilon_{Q}(n_{u},n_{d},n_{s}=0)
   +\theta(r-R)\varepsilon_{H}(n_{n},n_{p},n_{B\neq N}=0) 
\nonumber \\ & &
   +\varepsilon_{L}[n_{l}(r)]+\varepsilon_{S}(r)
   +\frac{1}{8\pi}{\bf E}(r)^{2}\ .
\end{eqnarray}
Here $r$ is the distance from the center of the droplet, 
$\varepsilon_{Q}$, $\varepsilon_{H}$, and $\varepsilon_{L}$ are given
by Eqs.\ (57), (40), and (41), respectively, $\varepsilon_{S}(r)$ is 
the increase in energy density due to the quark and nucleon
distributions in the interfacial layer whose thickness we assume to be
much smaller than $R$, and {\bf E}$(r)$ is the electric field.  The 
baryon density for the inhomogeneous phase is given by
\begin{equation}
n_{b,D}(r)=\theta(R-r)n_{b,Q}+\theta(r-R)n_{b,H}\ ,
\end{equation}
where $n_{b,Q}$ is the baryon density inside the droplet determined by
Eq.\ (58).  We thus obtain the potential energy by integrating the 
difference in the thermodynamic potential per unit volume at chemical 
potential $\mu_{b,H}$ between the initial metastable phase and the
inhomogeneous phase over the system volume $V$, which is related to
$P$ via $P=-\partial(V\varepsilon_{{\rm tot},H})/\partial V$, as
\begin{eqnarray}
   U(R)&=&\int_{V}dV
     \{\varepsilon_{D}(r)-\varepsilon_{{\rm tot},H}
     -\mu_{b,H}[n_{b,D}(r)-n_{b,H}]\}
\nonumber \\ &=&
   \frac{4\pi R^{3}}{3}n_{b,Q}(\mu_{b,Q}-\mu_{b,H})
   +\Delta E_{L}(R)+4\pi\sigma_{s} R^{2}+E_{C}(R)\ ,
\end{eqnarray}
where
\begin{equation}
\mu_{b,Q}=\frac{\varepsilon_{Q}(n_{u},n_{d},n_{s}=0)+
\varepsilon_{L}(n_{l,H})+P}{n_{b,Q}}
\end{equation}
is the chemical potential for the quark phase and
\begin{equation}
\Delta E_{L}(R)=\int_{V}dV\{\varepsilon_{L}[n_{l}(r)]
-\varepsilon_{L}(n_{l,H})\}
\end{equation}
is the excess of the lepton energy over the initial value.  Here 
$n_{l,H}$ is the number density of $l$ leptons in the initial
metastable phase of nuclear matter, $\sigma_{s}$ is the surface
tension, and $E_{C}$ is the electrostatic energy.

Since the quark-hadron interfacial properties are poorly known at
finite densities, the expression for $\sigma_{s}$ has been taken from 
the Fermi-gas model for a quark matter droplet in vacuum [33] as
\begin{eqnarray}
\sigma_{s}&=&\frac{3}{4\pi}\sum_{q   %=u,d,s
}\left\{\frac{\mu_{q}(\mu_{q}^{2}-m_{q}^{2})}{6}
-\frac{m_{q}^{2}(\mu_{q}-m_{q})}{3}-\frac{1}{3\pi}
\left[\mu_{q}^{3}\tan^{-1}\frac{(\mu_{q}^{2}-m_{q}^{2})^{1/2}}{m_{q}}
\right.\right.
\nonumber \\ & &
\left.\left.
-2\mu_{q}m_{q}(\mu_{q}^{2}-m_{q}^{2})^{1/2}
+m_{q}^{3}\ln\frac{\mu_{q}+(\mu_{q}^{2}-m_{q}^{2})^{1/2}}{m_{q}}
\right]\right\}\ ,
\end{eqnarray}
%\begin{equation}
%\sigma_{s}=\frac{3}{4\pi^{2}}\sum_{q   %=u,d,s
%}m_{q}\mu_{q}^{2}
%\end{equation} for $\mu_{q}\gg m_{q}$ and 
where $\alpha_{s}$ is set to be zero.  
This expression arises from the reduction in
quark density of states due to the presence of the droplet surface.
The typical value of $\sigma_{s}$ for matter without strangeness
is obtained as $\sigma_{s}\approx(3/4\pi^{2})(m_{u}\mu_{u}^{2}+
m_{d}\mu_{d}^{2})\sim 10$ MeV fm$^{-2}$, since 
$m_{u}\sim m_{d}\sim 10$ MeV and $\mu_{u}\sim\mu_{d}\sim 500$ MeV\@.
The foregoing Fermi-gas 
description includes neither $O(\alpha_{s})$ corrections nor curvature 
corrections.  The former have yet to be determined, while the latter 
destabilize a quark matter droplet in vacuum by producing an energy
\begin{equation}
E_{\rm curv}=\frac{R}{\pi}\sum_{q    %=u,d,s
}\mu_{q}^{2}
\end{equation}
for $\mu_{q}\gg m_{q}$ and $\alpha_{s}=0$ [34].  Instead of explicitly 
including these curvature corrections in the potential (68), we set
the range of $\sigma_{s}$ as 10 MeV fm$^{-2}$ 
$\leq\sigma_{s}\leq 50$ MeV fm$^{-2}$ by incorporating
%in such a way as to incorporate 
into $\sigma_{s}$ the increase in the 
interfacial energy yielded by the curvature energy (72) of the droplet
having a radius in excess of $\sim 3$ fm.  Shell effects as
encountered in nuclei are likely to appear remarkably in light quark
matter droplets having a baryon number $\lesssim 10$ [28], which cannot
be well described by the Fermi-gas model for finite quark matter as
adopted here.  Nevertheless, we may assume this model to be useful, 
since the baryon number contained in a virtual droplet moving under
the potential barrier is $\gtrsim 10$ (typically of order 100).  
The contribution of nucleons to the interfacial energy, ignored here, 
is expected to be much smaller than that of quarks at 
$n_{b,H}\sim n_{0}$, but it is uncertain in the considered density 
range $n_{b,H}\sim 4$--$6n_{0}$.

In the absence of the lepton screening on the droplet charge,
\begin{equation}
Z_{0}e=\frac{4\pi R^{3}\rho_{Q0}}{3}\ ,
\end{equation}
with 
\begin{equation}
%\rho_{Q0}=e\left(\sum_{q}q_{q}n_{q}-n_{e,H}-n_{\mu,H}\right)\ ,
\rho_{Q0}=e\left(\frac{2n_{u}-n_{d}-n_{s}}{3}-n_{e,H}-n_{\mu,H}\right)\ ,
\end{equation}
we obtain $E_{C}=3Z_{0}^{2}e^{2}/5R$ and 
\begin{equation}
 \Delta E_{L}=Z_{0}\mu_{e,H}\ , 
\end{equation}
where $\mu_{e,H}$ is the chemical potential of electrons in the
initial metastable phase of nuclear matter; typically, 
$\mu_{e,H}\sim 250$--300 MeV\@.  Here Eq.\ (75) reflects the fact that
the global charge neutrality in the system volume $V$ is guaranteed by
the lepton gas which satisfies the $\beta$-equilibrium condition (51).
The excess lepton energy $\Delta E_{L}$, which is proportional to 
$R^{3}$, is naturally absorbed into the chemical-potential difference 
term in the potential (68).  The number densities $n_{u}$ and $n_{d}$
are then determined from the flavor conservation (61) and the pressure
equilibrium between quark and nuclear matter
\begin{equation}
   P_{Q}=P_{H}-\frac{\Delta E_{L}}{4\pi R^{3}}\ ,
\end{equation}
where $P_{Q}$ and $P_{H}$ are given by Eqs.\ (60) and (43),
respectively.  In Eq.\ (76) the surface and Coulomb pressures arising 
from $4\pi\sigma_{s}R^{2}$ and $E_{C}(R)$ have been ignored since
these pressures are much smaller than the bulk pressures $P_{Q}$ and 
$P_{H}$.\footnote{As long as $\sigma_{s}\propto n_{q}^{2/3}$, the 
surface pressure vanishes.} 
$n_{u}$ and $n_{d}$ are thus independent of $R$.  It is instructive to 
note that the flavor conservation (61) and the charge neutrality (49)
in the hadronic phase result in the proportionality of $\rho_{Q0}$ to
$n_{e,H}+n_{\mu,H}$ and to $n_{b,Q}/n_{b,H}-1$: 
\begin{equation}
 \rho_{Q0}=e(n_{e,H}+n_{\mu,H})
 \left(\frac{n_{b,Q}}{n_{b,H}}-1\right)\ ,
\end{equation}
where the ratio $n_{b,Q}/n_{b,H}$ amounts to 1.2--1.8.

We next consider the screening action by leptons on the excess droplet 
charge $Z_{0}e$.  The number density $n_{l}(r)$ of $l$ leptons, whose 
deviation $\delta n_{l}(r)$ from $n_{l,H}$ is small enough to be
treated as a perturbation as will be shown below, has been determined 
from the linear Thomas-Fermi screening theory [35].  This
determination [11] leads to
\begin{equation}
  \delta n_{l}(r)=\left.\frac{\partial n_{l}}{\partial\mu_{l}}
\right|_{n_{l}=n_{l,H}}e\phi(r)
\equiv \frac{\kappa_{l}^{2}}{4\pi e^{2}}e\phi(r)\ ,
\end{equation}
where $\phi(r)$ is the electrostatic potential derived from the
Poisson equation 
$\nabla^{2}\phi(r)-\kappa_{L}^{2}\phi(r)=-4\pi\rho_{Q0}\theta(R-r)$
with the reciprocal of the screening length 
$\kappa_{L}=\sqrt{\kappa_{e}^{2}+\kappa_{\mu}^{2}}$.
%$\kappa_{L}=\sqrt{\lambda_{{\rm TF},e}^{-2}+\lambda_{{\rm TF},\mu}^{-2}}$.
Its solution reads
\begin{equation}
  \phi(r)=\left\{ \begin{array}{lll}
       \displaystyle{\frac{4\pi\rho_{Q0}}{\kappa_{L}^{2}}}
       \left[1-\exp(-\kappa_{L} R)(1+\kappa_{L} R)
       \displaystyle{\frac{\sinh(\kappa_{L} r)}{\kappa_{L} r}}\right]\ ,
         & \mbox{$r\leq R\ ,$} \\
             \\
       \displaystyle{\frac{4\pi\rho_{Q0}}{\kappa_{L}^{2}}}
        [\kappa_{L} R\cosh(\kappa_{L} R)-\sinh(\kappa_{L} R)]
       \displaystyle{\frac{\exp(-\kappa_{L} r)}{\kappa_{L} r}}\ ,
         & \mbox{$r>R\ .$}
 \end{array} \right.
\end{equation}
The resulting electrostatic energy is given by
\begin{equation}
  E_{C}(R)=\frac{2\pi^{2}\rho_{Q0}^{2}}{\kappa_{L}^{5}}
  \{-3+(\kappa_{L} R)^{2}+\exp(-2\kappa_{L} R)[3+6\kappa_{L} R
  +5(\kappa_{L} R)^{2}
   +2(\kappa_{L} R)^{3}]\}\ .
\end{equation}
We have confirmed that the linear Thomas-Fermi approximation may be 
safely used; the validating condition $\mu_{e,H}\gg e\phi(r)$ is well 
satisfied, since $e\phi(r)/\mu_{e,H}\lesssim 0.1$ for $R\lesssim 10$
fm.\footnote{The energy increment yielded by the gradient of 
$\delta n_{l}$, present over a scale of the screening length
$\kappa_{L}^{-1}\sim5$ fm inward and outward from the droplet surface,
is neglected in the present study.}

Figure 6(a) exhibits the potential for the formation and growth of a 
quark matter droplet, characterized by $b=100$ MeV fm$^{-3}$, 
$\alpha_{s}=0.4$, and $\sigma_{s}=30$ MeV fm$^{-2}$, in the initial 
metastable phase of nuclear matter at pressures above 
$P_{0}\approx 453$ MeV fm$^{-3}$.  This figure contains three cases in 
which the electrostatic energy is set to be zero, the nonscreened
value $3Z_{0}^{2}e^{2}/5R$, and the lepton-screened value given by 
Eq.\ (80).  By comparing these cases, we observe that the
lepton-screened case shows no dominance of the Coulomb energy over
$U(R)$ for large $R$ in contrast to the nonscreened case, and agrees 
fairly well with the case in which $E_{C}=0$.  This is because the
cloud of leptons spreads 
inside the droplet and from its surface outward
%from the droplet center 
over a scale of the 
screening length $\kappa_{L}^{-1}\sim 5$ fm 
(see Fig.\ 9 in the Appendix).  
We have thus confirmed a salient feature found in Ref.\ [11] that 
the lepton screening effects prevent the formation of a Coulomb barrier 
which causes the droplet to remain finite.  The good agreement between 
the lepton-screened and $E_{C}=0$ cases allows us to calculate the 
nucleation time without including $E_{C}$ in the potential (68).
Hereafter, we shall leave out $E_{C}$ in the potential (68); the 
resulting potential is characterized by the critical droplet radius
\begin{equation}
R_{c}=\frac{3\sigma_{s}}{n_{b,Q}(\mu_{b,H}-\mu_{b,Q})
-\rho_{Q0}\mu_{e,H}/e}
\end{equation}
and by the barrier height $U_{0}=(4/27)4\pi\sigma_{s} R_{c}^{2}$.

The effective mass $M(R)$ of a quark matter droplet in $\beta$-stable 
nuclear matter may be estimated from the kinetic energies of nucleons 
and leptons, in a way analogous to that described in Sec.\ II A\@.  We 
obtain the nucleon kinetic energy $K_{H}|_{n_{B\neq N}=0}$ of the form 
(3) by deriving the baryon velocity field from the baryon continuity 
equation and the boundary condition at the droplet surface;
the expression for $K_{H}$ is
\begin{equation}
K_{H}=2\pi\varepsilon_{H}(n_{n},n_{p},n_{B\neq N})
\left(1-\frac{n_{b,Q}}{n_{b,H}}\right)^{2}R^{3}{\dot R}^{2}\ .
\end{equation}
By substituting the lepton distribution (78) into the lepton
continuity equation, the lepton velocity field can be determined as
\begin{eqnarray}
  v_{l}(r) &=& -\frac{\rho_{Q0}}{n_{l,H}e}
\left(\frac{\kappa_{l}}{\kappa_{L}}\right)^{2}
\frac{{\dot R}}{(\kappa_{L} r)^{2}}
   \nonumber \\ & &
   \nonumber \\ & &
\times\left\{ \begin{array}{lll}
  \kappa_{L} R\exp(-\kappa_{L} R)[\kappa_{L}r\cosh(\kappa_{L} r) 
  -\sinh(\kappa_{L} r)]\ ,
         & \mbox{$r\leq R\ ,$} \\
             \\
%  \kappa_{L} R\exp(-\kappa_{L} R)[\cosh(\kappa_{L} r)\kappa_{L} 
%  r-\sinh(\kappa_{L} r)]\ ,
 (\kappa_{L} R)^{2}-\kappa_{L}R\sinh(\kappa_{L} R)
                   (1+\kappa_{L} r)\exp(-\kappa_{L} r)\ ,
         & \mbox{$r>R\ .$}
 \end{array} \right.
\end{eqnarray}
The lepton kinetic energy $K_{L}$, which is evaluated from Eq.\ (83)
 as
\begin{equation}
K_{L}< 2\pi\varepsilon_{L}(n_{l,H})
  \left[\sum_{l}\frac{\kappa_{l}^{4}}{\kappa_{L}^{4}}
%K_{L}<\sum_{l}\frac{2\pi}{\kappa_{L}^{4}\lambda_{{\rm TF},l}^{4}}
  \left(\frac{\rho_{Q0}}{n_{l,H}e}\right)^{2}\right]
 R^{3}{\dot R}^{2}\ ,
\end{equation}
proves negligible, partly because
$\varepsilon_{H}(n_{n},n_{p},n_{B\neq N}=0)\sim 1000$ MeV fm$^{-3}$ is
much larger than $\varepsilon_{L}(n_{l,H})$ of order 10 MeV fm$^{-3}$
and partly because Eq.\ (77) and $n_{e,H}\sim n_{\mu,H}$ 
ensure $(n_{b,Q}/n_{b,H}-1)^{2}\sim (\kappa_{l}^{4}/\kappa_{L}^{4})
(\rho_{Q0}/n_{l,H}e)^{2}$.
The effective mass $M(R)$ is finally obtained as
\begin{equation}
   M(R)=4\pi\varepsilon_{H}(n_{n},n_{p},n_{B\neq N}=0)
\left(1-\frac{n_{b,Q}}{n_{b,H}}\right)^{2}R^{3}\ .
\end{equation}

The effect of energy dissipation discussed in Sec.\ II C plays a 
crucial role in determining the time needed to form a quark matter 
droplet in $\beta$-equilibrated nuclear matter, 
as shown in Ref.\ [12].  Let us now assume that the nucleons 
and the leptons are in a normal fluid state.  Then, 
elementary excitations of particle species $i$ ($i=N,l$) appear as 
quasiparticles in the vicinity of their respective Fermi surfaces 
during a fluctuation of $R$.  
Since the mean free path of $i$ excitations is estimated\footnote{
This estimate of $l_{i}$ is based on the Landau theory of the Fermi 
liquids [36].  Here we have assumed that the coefficient $C_{i}$
affixed to $n_{i}^{-1/3}(\mu_{i}-\mu_{i}|_{n_{i}=0})^{2}T^{-2}$ 
is of order unity.  However, $C_{i}$ depends on the interaction 
between quasiparticles (i.e., the Landau parameters).
For nucleons, such an interaction is repulsive in the high
density region of interest here [37,38]; the values of $C_{N}$ range
$\sim0.1$--1.  For leptons, being relativistic and degenerate,
polarization effects of the medium on the exchanged photons 
[27,39] induce attraction between quasiparticles;
the values of $C_{l}$ are roughly 10. These considerations
of $C_{i}$ do not change the conclusion $l_{i}\gg R_{c}$.} 
as 
$l_{i}\sim n_{i}^{-1/3}(\mu_{i}-\mu_{i}|_{n_{i}=0})^{2}T^{-2}>10^{2}$ 
fm, we can observe that each $l_{i}$ is sufficiently large compared
with the critical droplet radius $R_{c}$ ranging typically 1--5 fm. 
Consequently, the nucleon and lepton excitations behave 
ballistically and collide with the droplet surface. The resulting 
transfer of momentum flux exerts on the droplet an Ohmic friction 
force of the form analogous to Eq.\ (22); this is written
as a sum of nucleonic and leptonic terms 
\begin{equation}
F=F_{H}|_{n_{B\neq N}=0}+F_{L}\ ,
\end{equation}
with 
\begin{equation}
F_{H}=-16\pi\left(\frac{n_{b,Q}}{n_{b,H}}-1\right)^{2}
\left(\sum_{B}\alpha_{B}n_{B}k_{F,B}\right)R^{2}{\dot R}\ ,
\end{equation}
\begin{equation}
F_{L}=-16\pi f(\kappa_{L} R) \left[\sum_{l}
\frac{\kappa_{l}^{4}}{\kappa_{L}^{4}}
%\frac{1}{\kappa_{L}^{4}\lambda_{{\rm TF},l}^{4}}
\left(\frac{\rho_{Q0}}{n_{l,H}e}\right)^{2}
%\left(\sum_{l}
\alpha_{l}n_{l,H}k_{F,l,H}\right]
R^{2}{\dot R}\ .
\end{equation}
Here $k_{F,l,H}=(3\pi^{2}n_{l,H})^{1/3}$ is the Fermi momentum of $l$ 
leptons in the metastable phase, $\alpha_{i}$ is a factor of order
unity that depends on the properties of $i$ excitations and on their 
interactions with the droplet surface, and $f(\kappa_{L} R)$ is a 
function that stems from the velocity field (83) occurring due to the 
lepton screening effects, behaves as $\propto (\kappa_{L} R)^{3}$ for 
$\kappa_{L} R\ll 1$, and monotonically approaches unity as 
$\kappa_{L} R$ increases.\footnote{Note that Eq.\ (86) reduces to 
Eq.\ (8) in Ref.\ [12] in the absence of muons and in the limit of 
$\kappa_{L} R\rightarrow\infty$.  Thus, the contribution of the 
electron quasiparticles to the friction force was rather
overestimated in Ref.\ [12].}  Since 
$\sum_{N}n_{N}k_{F,N}\sim 250$--450 MeV fm$^{-3}$ is one order of 
magnitude larger than 
$\sum_{l}n_{l,H}k_{F,l,H}\sim 30$--70 MeV fm$^{-3}$, 
we ignore the leptonic contribution (88) to
the Ohmic friction force.  The Fermi velocity 
$v_{F,H}|_{n_{B\neq N}=0}$ averaged over nucleon species, where
$v_{F,H}$ is defined as
\begin{equation}
v_{F,H}=\frac{\sum_{B}n_{B}k_{F,B}}
        {\varepsilon_{H}(n_{n},n_{p},n_{B\neq N})}\ ,
\end{equation}
is $\sim0.4c$, and hence $v_{F,H}|_{n_{B\neq N}=0}$ is in excess of 
$\sqrt{2U_{0}/M(R_{c})}\sim0.1$--0.3$c$.  As a consequence, by
recalling how to distinguish between weak and strong dissipation as 
described in Sec.\ II C, we find that except for $\alpha_{N}\ll 1$,
the nucleation time is determined by the dissipative processes rather 
than by the reversible droplet motion.

The effects of relativity and dynamical compressibility as considered
in Secs.\ II B and II D make only a negligible change in the estimates
of the time necessary for the quantum nucleation of quark matter in
the metastable phase of nuclear matter.  This is because the droplet 
surface moves under the potential barrier at a velocity [typically 
$\sim\sqrt{2U_{0}/M(R_{c})}$] fairly low compared with $c_{s}\sim0.8c$
(nuclear matter) and $c_{s}\sim0.6c$ (deconfined matter).  
%Thus, these considerations 
These comparisons of velocity scales lead 
%roughly 
us to the effective action of a form similar to Eq.\ (24):
\begin{eqnarray}
   S_{\rm eff}[R(\tau')] &=& \int_{-\beta/2}^{\beta/2}d\tau'
   \left\{
% \frac{1}{2}M(R)\dot{R}^{2}}+
    \frac{}{} U(R)
    +\alpha\left(1-\frac{n_{b,Q}}{n_{b,H}}\right)^{2}
    \left(\sum_{N}n_{N}k_{F,N}\right) \right. 
\nonumber \\ & & \left.
   \times\int_{-\beta/2}^{\beta/2}d\tau''
    (R_{\tau'}^{2}-R_{\tau''}^{2})^{2}\frac{(\pi/\beta)^{2}}
    {\sin^{2}\pi (\tau'-\tau'')/\beta}
    \right\}\ .
\end{eqnarray}
Here $U(R)$ is taken from Eq.\ (68) in which we set $E_{C}=0$, and  
$\alpha_{n}=\alpha_{p}\equiv\alpha$ is assumed for simplicity.  The 
extremal action $A(T)$ for the quantum nucleation is obtained in a
form identical to Eq.\ (29) as
\begin{equation}
A(T)=4\pi\alpha\left(1-\frac{n_{b,Q}}{n_{b,H}}\right)^{2}
     \left(\sum_{N}n_{N}k_{F,N}\right)R_{c}^{4}s(T)\ ,
\end{equation}
where $R_{c}$ is given by Eq.\ (81).

%\vspace*{0.5cm}
%\begin{center}
%2. Hyperonic matter
%\end{center}
%\vspace*{0.3cm}
\subsubsection{Hyperonic matter}
\label{sec:hm}

Let us now construct the effective action for the development of a
quark matter droplet in the initial metastable phase of 
$\beta$-equilibrated hadronic matter with hyperons.  In this
construction, we may utilize expressions (65)--(89) developed for
matter without hyperons by identifying the quantities characterizing 
the metastable phase at a given $P$, i.e., 
$\varepsilon_{{\rm tot},H}$, $\mu_{b,H}$, $\mu_{e,H}$, $n_{b,H}$, and 
$n_{l,H}$, with those for matter with hyperons as well as by removing 
the constraints $n_{B\neq N}=0$ and $n_{s}=0$.  Notice that the lepton 
fractions plotted in Fig.\ 2(b) are sufficiently small at baryon 
densities of interest, $n_{b,H}\sim$ 4--6$n_{0}$, to confine the ratio 
$|\rho_{Q0}/n_{b,Q}e|$ of the electric charge to the baryon number 
inside the droplet within $10^{-2}$.  This behavior stems from the 
fact that the ratio $|\rho_{Q0}/n_{b,Q}e|$ is proportional to the sum
of the lepton fractions $(n_{e,H}+n_{\mu,H})/n_{b,H}$, as can be seen 
from Eq.\ (77).  When using expressions (65)--(89), therefore, we omit 
the quantities associated with electricity by setting $\rho_{Q0}=0$. 
Correspondingly, the screening action by the charged components is 
neglected.  Such an omission is valid for the droplet sizes of
interest here, $R\lesssim20$ fm.  We have confirmed that for $R\gg20$ fm,
the Coulomb energy continues to be trivial due mainly to the screening
action by leptons and due partly to that by charged baryons.

By noting Eqs.\ (68) and (69), the potential $U(R)$ for the formation 
and growth of a quark matter droplet in $\beta$-stable hyperonic matter 
has been 
%thus 
written as 
\begin{equation}
 U(R)=\frac{4\pi R^{3}}{3}n_{b,Q}(\mu_{b,Q}-\mu_{b,H})
      +4\pi\sigma_{s} R^{2}\ ,
\end{equation}
with
\begin{equation}
\mu_{b,Q}=\frac{\varepsilon_{Q}(n_{q})+
\varepsilon_{L}(n_{l,H})+P}{n_{b,Q}}\ ,
\end{equation}
where $n_{b,Q}$, $\varepsilon_{Q}$, and $\varepsilon_{L}$ are given by 
Eqs.\ (58), (57), and (41).  Here $\sigma_{s}$, the surface tension,
has been assumed to range from 10 to 50 MeV fm$^{-2}$ by using the 
Fermi-gas expression (71); this range stems primarily from the 
uncertainty in $m_{s}$.  The curvature term (72) contributes only a 
little to the interfacial energy of the droplet with typical 
$R\sim5$--15 fm.  The quark number densities $n_{u}$, $n_{d}$, and 
$n_{s}$ have been obtained from the flavor conservation (61) and the 
pressure equilibrium between quarks and baryons $P_{Q}=P_{H}$, where 
$P_{Q}$ and $P_{H}$ are given by Eqs.\ (60) and (43), respectively;
modification of such pressure equilibrium by the interfacial energy 
is negligibly small for typical $R$.  In Fig.\ 6(b) we have depicted
the potential energy (92),
%of a quark matter droplet in the initial 
%metastable phase of hyperonic matter, 
calculated for $b=100$ MeV fm$^{-3}$, $\alpha_{s}=0.4$, $m_{s}=0$,
$\sigma_{s}=30$ MeV fm$^{-2}$, and
various pressures above $P_{0}\approx 163$ MeV fm$^{-3}$.  The form of 
this potential is identical to the standard one denoted by Eq.\ (1);
the critical droplet radius is given by
\begin{equation}
R_{c}=\frac{3\sigma_{s}}{n_{b,Q}(\mu_{b,H}-\mu_{b,Q})}\ ,
\end{equation}
from which the potential barrier height $U_{0}$ is determined as 
$U_{0}=(4/27)4\pi\sigma_{s} R_{c}^{2}$.

The effective mass $M(R)$ of a quark matter droplet in the metastable
phase of hyperonic matter is then obtained from the baryon kinetic 
energy $K_{H}$ given by Eq.\ (82) as
\begin{equation}
   M(R)=4\pi\varepsilon_{H}(n_{B})
\left(1-\frac{n_{b,Q}}{n_{b,H}}\right)^{2}R^{3}\ .
\end{equation}
In this determination we ignore the lepton and quark kinetic energies 
since we have set $\rho_{Q0}=0$.
%since $\rho_{Q0}\approx 0$.  
Expression (95) does not include the 
relativistic effect on the hydrodynamic mass flow of baryons. 
As will be shown in the next paragraph, the baryon velocity field,
given by Eq.\ (2) in which $n_{1}=n_{b,H}$ and $n_{2}=n_{b,Q}$, 
is far smaller than $c$, although the droplet motion itself is 
relativistic in many cases.

%In setting the structure of 
In constructing the effective action for the system having
a quark matter droplet in $\beta$-equilibrated hyperonic matter, 
it is instructive to 
compare the velocities $|{\dot R}|$, $c_{s}$, and $v_{F,H}$
characterizing the droplet motion under the potential barrier.  It is
of great importance to note that over a wide range of the parameters 
$\sigma_{s}$, $b$, $\alpha_{s}$, and $m_{s}$ with relatively large 
$\alpha_{s}$ ($\gtrsim0.2$) and $\sigma_{s}$ ($\gtrsim 30$ MeV fm$^{-2}$),
the density discontinuity $n_{b,Q}-n_{b,H}$ is small enough to make 
$\sqrt{2U_{0}/M(R_{c})}$ higher than 0.3$c$ for typical 
$R_{c}\sim5$--15 fm.  We should, therefore, take into account the
effect of relativity on the quantum nucleation as mentioned in 
Sec.\ II B\@.  The velocity of sound, $c_{s}\sim0.5c$, 
of hyperonic matter cannot fully surpass the rate $|{\dot R}|$ of
growth of a virtual droplet ranging typically $\sim0.1$--$1c$.  Thus,
the effect of the dynamical compressibility of the metastable phase as 
discussed in Sec.\ II D is likely to have consequence to the quantum 
nucleation at temperatures near the quantum-thermal 
crossover point.\footnote{In the present analysis, 
no attention is paid to the dynamical compressibility of the
deconfined phase, which may affect the kinetic term included in the 
effective action through a change in the boundary condition at the 
droplet surface having the form (32).}  The absolute value of the
baryon velocity field, which is less than 
$|(1-n_{b,Q}/n_{b,H}){\dot R}|$ [see Eq.\ (2)], is limited within
$0.1c$.  This leads us to use the effective mass $M(R)$ 
given by Eq.\ (95).  The effect of 
energy dissipation is controlled by low-lying excitations of baryons 
induced by the fluctuation of $R$.  These excitations, whose mean free
paths are sufficiently large compared with typical $R_{c}$
(except for minor components), collide with the droplet surface 
and exert the Ohmic friction force $F_{H}$ given by Eq.\ (87) 
on the droplet.  Here we have assumed the baryons to be
in a normal fluid state, and we have ignored the negligible 
contributions of leptons and quarks to the energy dissipation.  The
Fermi velocity $v_{F,H}$ given by Eq.\ (89), averaged over baryon
species, is estimated to be $\sim0.3c$, so that $|{\dot R}|>v_{F,H}$
is satisfied in the relativistic regime.  We thus confine ourselves to
%considerations of 
the dissipationless cases.  We finally write the effective action as
\begin{eqnarray}
   S_{\rm eff}[R(\tau')]&=&\int_{-\beta/2}^{\beta/2}d\tau'
   \left\{
   \left(-1+\sqrt{1+{\dot{R}}^{2}}\right)
   \frac{1}{{\dot R}^{2}}
%   \left[1-\sqrt{1-\left(\frac{\dot{R}}{c}\right)^{2}}\right]
%   \left(\frac{c}{{\dot R}}\right)^{2}
   \right.
\nonumber \\ & &
   \left.
   \times\int_{R(\tau')}^{\infty}dr 4\pi\varepsilon_{H}(n_{B})r^{2}
   \left[\left(\frac{\partial\varphi}{\partial r}\right)^{2}
         +\frac{1}{c_{s}^{2}}
   \left(\frac{\partial\varphi}{\partial\tau'}\right)^{2}\right]
     +U(R)
   \right\}\ .
\end{eqnarray}
Here $U(R)$ is given by Eq.\ (92), and $\varphi(r)$ is the velocity 
potential derived from the wave equation (31) and the boundary
condition (32) in which $n_{1}=n_{b,H}$ and $n_{2}=n_{b,Q}$.

%\vspace*{0.5cm}
%\begin{center}
%{\bf B. Nucleation time}   % NOTE T and P dependence.
%\end{center}
%\vspace*{0.3cm}
\subsection{Nucleation time}
\label{sec:nt}

Let us now proceed to evaluate the time $\tau$ needed to form a real 
quark matter droplet via quantum tunneling in the metastable phase of 
$\beta$-equilibrated hadronic matter with (without) hyperons, in the 
regime of overpressures and relatively low temperatures, by using the 
effective action (90) [(96)].  We may obtain $\tau$ within the 
exponential accuracy as 
\begin{equation}
  \tau=\nu_{0}^{-1}\exp(A)\ ,
\end{equation}
where $\nu_{0}$ is the frequency of the zero-point oscillation around 
$R=0$, and $A$ is the corresponding extremal value of the effective 
action $S_{\rm eff}$.  Here we have ignored the thermal effect on the
pre-exponential factor of $\tau$, since the frequencies of the higher 
order oscillations are logarithmically as large as $\nu_{0}$.  It may
also be assumed that the energy dissipation and dynamical 
compressibility in hadronic matter have no influence on the 
pre-exponential factor, 
because the velocities $v_{F,H}$ and $c_{s}$ take on nearly the same 
order of magnitude
as the rate $|{\dot R}|$ of growth of a virtual droplet.
The values of $\nu_{0}^{-1}$ obtained by using Eq.\ (8) amounts to 
$\sim10^{-23}$--10$^{-22}$ s, being the time scale for strong 
interactions, so we set $\nu_{0}^{-1}=10^{-23}$ s.

In the absence of hyperons, we recall that the extremal effective
action for the quantum nucleation may be roughly expressed as Eq.\
(91).  Since the quantity $s(T)$ included in Eq.\ (29) depends only
weakly on temperatures up to the quantum-thermal crossover value that 
will be found to be $\lesssim10$ MeV, we may estimate the time $\tau$ 
required to form a single quark matter droplet by 
setting the temperature to be zero.
%ignoring the effects of finite temperature on the quantum nucleation.
The zero-temperature 
values of $\tau$ obtained for $b=100$ MeV fm$^{-3}$, $\alpha_{s}=0.4$, 
$m_{s}=0$, $\sigma_{s}=10,30,50$ MeV fm$^{-2}$, and $\alpha=1$ have
been depicted in Fig.\ 7(a) as a function of $P$.  With the increment 
in $P$ from the pressure, $P_{0}\cong453$ MeV fm$^{-3}$, of the static 
deconfinement transition, the critical droplet radius $R_{c}$ decreases 
as Fig.\ 6(a) exhibits.  Such behavior leads to a decrease in $A$ 
$(\propto R_{c}^{4})$ and hence to an exponential reduction of $\tau$.
This indicates that at fixed $P$, $\tau$ decreases exponentially as
$\sigma_{s}(\propto R_{c})$ decreases linearly.
As a consequence, the ambiguity in $\sigma_{s}$ disperses the 
overpressure $\Delta P=P-P_{0}$ between $\sim30$ and $\sim140$ 
MeV fm$^{-3}$ at a typical $\tau$.  For comparison, 
the formation time $\tau$ in the dissipationless case ($\alpha=0$) has 
also been evaluated by using the formalism written in Sec.\ II A; the 
results have been plotted in Fig.\ 7(a).  The role of energy
dissipation in increasing the overpressure $\Delta P$ needed to give
a constant $\tau$ has been thus clarified.

In the case in which the hadronic phase contains hyperons, we may 
roughly estimate the extremal effective action by using Eq.\ (96). 
Recall that the extremal value of the compressibility term $S_{C}$ 
is likely to behave as $\propto (T/T_{0})^{4}$ for $T\ll T_{0}$, as 
stated in Sec.\ II D\@.  Here $T_{0}$ is the quantum-thermal crossover 
temperature found from Eq.\ (26) as $A(E_{0})=U_{0}/T_{0}$, where 
$A(E_{0})$ is obtained from Eq.\ (18) at a given $P$ and $T=0$.  Since 
the value of $T_{0}$ that will be estimated to be $\gtrsim10$ MeV is far 
larger than the temperature of matter in the neutron star core
$T_{\rm NS}\lesssim0.1$ MeV, we ignore the contribution of the dynamical
compressibility of the hadronic phase to the extremal effective action. 
Thermal effects excite a droplet oscillating around $R=0$ from the 
ground state of energy $E_{0}$ to the $m$th bound state of energy 
$E_{m}$ according to the Boltzmann distribution, and hence produce a 
probability of quantum-mechanical formation of a real droplet having
the energy $E_{m}$, relative to the ground-state tunneling
probability, as
\begin{equation}
p_{m}=\exp\left(-\frac{E_{m}-E_{0}}{T}\right)
%\frac{\exp\left(-\frac{E_{m}-E_{0}}{T}\right)}
%    {\sum_{m}\exp\left(-\frac{E_{m}-E_{0}}{T}\right)}
    \exp[A(E_{0})-A(E_{m})]\ . 
\end{equation}
Here $m=1,2,3,\ldots$, and $E_{m}$ is determined from
$I(E_{m})=2\pi(m+m_{0}+3/4)$, where $I(E)$ and $m_{0}$ are given by 
Eqs.\ (16) and (17), respectively.  At typical 
pressures, the difference $E_{m}-E_{m-1}$ is of order 100 MeV, whereas 
$A(E_{m-1})-A(E_{m})$ takes on a value of order $10$.  At 
$T=T_{\rm NS}$, the probability $p_{m}$ 
%of formation of a droplet being in the $m$-th excited state 
is virtually zero, so we take no 
account of the thermal enhancement of the tunneling probability.
% NOTE THAT LK ANALYSIS LEADS TO ABSENCE OF
% THERMALLY ASSISTED TUNNELING BY ASSUMING CONTINUOUS ENERGY SPECTRUM.

The nucleation time $\tau$ has been then calculated at zero
temperature for hadronic matter with hyperons by substituting the 
potential (92) and the effective mass (95) into Eq.\ (18).  In 
Fig.\ 7(b) we have plotted the results evaluated for 
$b=100$ MeV fm$^{-3}$, $\alpha_{s}=0.4$, $m_{s}=0$, and 
$\sigma_{s}=10,30,50$ MeV fm$^{-2}$, 
together with the nonrelativistic results obtained from Eq.\ (10).
With increasing $P$ from the pressure, $P_{0}\cong163$ MeV fm$^{-3}$,
of the static deconfinement transition,
the potential barrier is lowered and narrowed for fixed $\sigma_{s}$, 
as illustrated in Fig.\ 6(b).  The time $\tau$ thus shows an 
exponential $P$ dependence in the overpressure regime, as Fig.\ 7(b) 
displays.  We see by comparison that for large $\sigma_{s}$, the
effect of relativity manifests itself as an exponential reduction of 
$\tau$ at fixed $P$.  This feature reflects the fact that the lower 
bound $E_{\rm min}$ of the positive energy region, given by the
maximum value of the quantity $U(R)-2M(R)$ behaving as Fig.\ 6(b)
exhibits, renders the ratio $E_{0}/U_{0}$ higher than its 
nonrelativistic value.  The larger $\sigma_{s}$, the more 
relativistically the droplet moves under the potential barrier. 
Accordingly, the extent of the reduction in $\tau$ develops at
constant $P$, as can be seen from Fig.\ 7(b).  For small $\sigma_{s}$, 
on the other hand, the nonrelativistic results for $\tau$ are slightly
smaller than the relativistic results.  This is because relativistic 
corrections act to enhance the kinetic energy of the droplet, as is 
evident from Eq.\ (12).  We also observe in Fig.\ 7(b) that due to the
uncertainty in $\sigma_{s}$, $\Delta P$ spreads between $\sim10$ and 
$\sim60$ MeV fm$^{-3}$ at a typical $\tau$.

%\vspace*{0.5cm}
%\begin{center}
%{\bf C. Stellar conditions}    
%\end{center}
%\vspace*{0.3cm}
\subsection{Stellar conditions}
\label{sec:sc}

We turn to the evaluation of the time needed to form the first quark 
matter droplet in a neutron star whose core consists of $\beta$-stable 
hadronic matter with and without hyperons.  This evaluation may be
made according to $\tau/N_{s}$, where $N_{s}$ is the number of virtual
centers of droplet formation in the star, and $\tau$ is the formation 
time in the hadronic system including only a single 
nucleation center as obtained in the preceding subsection.  
%Let us assume 
Here we have assumed
the nucleation to be $homogeneous$: The star retains its 
hydrostatic equilibrium configuration during its evolution and
contains no impurities leading to the formation of a seed of quark 
matter.  $N_{s}$ can be estimated as
%We may thus estimate $N_{s}$ as
\begin{equation}
N_{s}\sim\frac{3V_{c}}{4\pi R_{c}^{3}}\ ,
\end{equation}
where $V_{c}$ is the volume of the central region in which the value of 
$\tau$ determined by its pressure remains within an order of magnitude 
of the value calculated at the star's center.  The volume $V_{c}$ has 
been obtained as $V_{c}\sim10^{6}$ m$^{3}$, irrespective of the
presence of hyperons, from the results for the pressure and density 
profiles of the star that were calculated in Sec.\ III from the 
Tolman-Oppenheimer-Volkoff equation and the equation of state of 
hadronic matter.  Hereafter, we set $N_{s}=10^{48}$ since $R_{c}$
ranges typically $\sim$2--15 fm.  The uncertainty in $N_{s}$, 
expected to be an order of magnitude or so, has little consequence.

In order that the first droplet may form during a time scale for the 
star's spin-down or accretion $\tau_{s}\sim$ yr--Gyr, the overpressure
should amount to such a value $\Delta P_{c}$ as to make $\tau/N_{s}$ 
comparable to $\tau_{s}$.  For example, we consider the case in which 
$b=100$ MeV fm$^{-3}$, $\alpha_{s}=0.4$, $m_{s}=0$, and $\alpha=1$ as 
exemplified in Fig.\ 7.  For hadronic matter with (without) hyperons, 
the critical overpressure $\Delta P_{c}$ lies between $\sim$10 
($\sim$30) and $\sim$60 ($\sim$140) MeV fm$^{-3}$ because of the 
uncertainty in the surface tension.  Correspondingly, the critical 
droplet radius $R_{c}$ and the potential barrier height $U_{0}$ take
on a value of $\sim$9--10 ($\sim$2.0--2.2) fm and 
$\sim$2--8$\times$10$^{3}$ ($\sim$100--400) MeV, respectively.

By using the barrier height $U_{0}$ thus evaluated, we have calculated
the quantum-thermal crossover temperature $T_{0}$ from the relation 
$A=U_{0}/T_{0}$; the results have been plotted in Fig.\ 8.  For 
comparison, we have also depicted the results obtained in the cases of 
strong dissipation $\alpha=0.2, 5$ and no dissipation $\alpha=0$ for 
nuclear matter as well as in the nonrelativistic case for hyperonic
matter.  We see from Fig.\ 8(a) that in the absence of hyperons, the 
crossover temperature $T_{0}$ is large compared with the temperature 
$T_{\rm NS}$ of matter in the stellar core, although the dissipation 
effects decrease $T_{0}$ considerably.  This result suggests that the 
formation of a quark matter droplet in the core composed of nuclear 
matter, if occurring, is likely to proceed from quantum fluctuations 
rather than from thermal fluctuations, as is consistent with the
result of Ref.\ [12].  Figure 8(b) shows that in the presence of 
hyperons, the obtained values of $T_{0}$ are 
%sufficiently 
again large
relative to $T_{\rm NS}$, and that the effect of relativity helps to 
enhance $T_{0}$ for large $\sigma_{s}$.  To be noted, however, is that 
the present estimate of $T_{0}$ ($\gtrsim 10$ MeV) is crude in itself.  
For such temperatures, the thermal effects may alter the
thermodynamic quantities included in the effective action (96), induce 
the heat conduction by frequent collisions between the excitations, 
enhance the rate of quantum-mechanical formation of a droplet by
raising its energy level around $R=0$ from the ground state to the 
low-lying excited states, and render the dynamical compressibility of 
the hadronic phase effective at increasing the formation rate.  At 
$T=T_{\rm NS}$, these effects are of little consequence, so it is safe
to conclude $T_{0}\gg T_{\rm NS}$.  We have also confirmed that in the
present calculations of $T_{0}$ allowing for and ignoring the presence 
of hyperons, the order of magnitude of $T_{0}$ is invariant over the
adopted range of the bag-model parameters.  We remark in passing that
a difference in $\tau/N_{s}$ by 10 orders of magnitude, expected from
the ambiguities in $\tau_{s}$, $N_{s}$, and $\nu_{0}$, alters 
$\Delta P_{c}$, $R_{c}$, and $T_{0}$ by less than 10\%.

We finally search for the range of the bag-model parameters $b$, 
$\alpha_{s}$, and $m_{s}$ in which the quantum-tunneling nucleation of
quark matter is expected to occur in the star during the compression 
of hadronic matter due to the stellar spin-down or accretion.
Let us now consider the critical condition of the bag-model parameters
for the presence of quark matter in the star: Initially, the star 
containing a hadronic core has a central pressure lower than 
$P_{\rm max}$, and subsequent compression, which increases the central 
pressure up to $P_{\rm max}$ in a time scale of $\tau_{s}$, drives the 
nucleation of quark matter in the central region.  Here we do not
inquire whether the resulting star continues to be gravitationally 
stable or not.  The combinations of $b$ and $\alpha_{s}$ have been 
then determined for hadronic matter with and without hyperons
in such a way as to satisfy the critical condition. 
In the absence of hyperons, we take $\alpha=1$ and 
$\sigma_{s}=30$ MeV fm$^{-2}$; in the presence of hyperons, 
$m_{s}$ and $\sigma_{s}$ have been set as $m_{s}=0, 150, 300$ MeV and
$\sigma_{s}=30$ MeV fm$^{-2}$.  The results have been plotted in 
Fig.\ 4.  In these four cases, the critical overpressure 
$\Delta P_{c}$ required to form the first droplet in the star results
in the shrinkage of the parameter range appropriate to the presence of
quark matter from that surrounded by the contour of 
$P_{0}=P_{\rm max}$.  We observe in the present calculations allowing 
for the presence of hyperons that a peak structure appears in the 
boundary of such a parameter range.  Along this boundary, the baryon 
density discontinuity $n_{b,Q}-n_{b,H}$ at deconfinement runs from 
$\sim-0.1$ to $\sim0.3$ fm$^{-3}$ with changing $b$ from 50 to 200 MeV 
fm$^{-3}$.  On the way, therefore, the point at which 
$n_{b,Q}=n_{b,H}$ does exist.  At this point, where the effective
droplet mass $M(R)$ given by Eq.\ (95) vanishes and so does the 
critical overpressure $\Delta P_{c}$, the boundary touches the contour
of $P_{0}=P_{\rm max}$, leading to the peak structure.  It is 
noteworthy that around the peak where the effective mass $M(R_{c})$ of 
the critical droplet is comparable to or smaller than the barrier
height $U_{0}$, the effect of relativity on the quantum nucleation 
becomes prominent.

%\newpage
%%\vspace*{0.5cm}
%\begin{center}
%{\bf V. CONCLUSIONS}
%\end{center}
%\vspace*{0.5cm}
\section{CONCLUSIONS}
\label{sec:conc}

We have examined the question of how the presence of hyperons in cold 
dense hadronic matter affects the dynamical properties of the 
first-order deconfinement transition that may occur in neutron stars. 
For this purpose, we have used the bag-model description of quark 
matter and the relativistic-mean-field description of hadronic matter.
The time needed to form a quark matter droplet has been then
calculated by using the quantum-tunneling nucleation theory which
allows for the electrostatic energy and the effects of relativity, 
energy dissipation, and finite compressibility.

In the calculations ignoring the presence of hyperons, we have found 
that the effective mass of a virtual droplet moving under the
potential barrier is sufficiently large to keep its growth rate 
fairly small compared with the Fermi velocity averaged
over nucleon species.  This results in an eminent property of the 
quantum-tunneling formation of a real droplet: The quantum tunneling
is controlled by the dissipative processes, in which the nucleon 
quasiparticles moving ballistically collide with the droplet surface, 
rather than by the reversible underbarrier motion of a droplet being
in a coherent state.  We have likewise found that inside a droplet of 
relatively small radius, an appreciable electric charge arises from
the pressure equilibrium between quarks and nucleons immersed in a 
roughly uniform sea of electrons and muons.  With increase in the 
droplet radius, the degenerate gas of electrons and muons becomes 
effective at screening the droplet charge, and at last prevents the 
Coulomb potential barrier from forming.  This result indicates that a 
real droplet, if appearing, would develop into bulk matter.  We note 
that these conclusions are consistent with the results of Refs.\ [11] 
and [12] obtained using a specific set of the bag-model parameters and
a rather simplified model for nuclear matter.

The estimates allowing for the presence of hyperons have shown that
the density discontinuity at deconfinement and the lepton fraction in 
the hadronic phase fall in magnitude considerably from the values 
obtained ignoring the presence of hyperons.  The resulting lepton 
fraction is small enough for the droplet charge to be nearly zero, 
allowing us to assume the charge neutrality everywhere in the system.
The decrease in the effective droplet mass due to the lessened density
discontinuity, when sufficient, drives the droplet to move 
relativistically under the potential barrier.  Not only does such a 
droplet motion cause the decelerating effect of energy dissipation on 
the droplet formation to decline, but also necessitates the 
relativistic treatment of the quantum nucleation theory.  A natural 
extension to the relativistic regime has been made, leading to an 
exponential increment in the tunneling probability.

Irrespective of whether hyperons exist or not, the finite 
compressibility of the hadronic phase is unlikely to have consequence 
in decreasing the kinetic energy of a droplet via sound emission at
low temperatures relevant to neutron star cores.  Such temperatures
have been found to be small compared with the temperature at which an 
abrupt transition from the quantum-tunneling to the thermal-activation 
nucleation takes place at a typical overpressure.

We have lastly investigated the range of the parameters $b$ and 
$\alpha_{s}$ where quark matter may 
%possibly 
appear in the core of a 
neutron star during the compression of hadronic matter due to the 
stellar spin-down or accretion.  It has been shown in the analysis of 
the static deconfinement transition that for the adopted models for
bulk hadronic matter and quark matter, the existence of hyperons and
the finiteness of $m_{s}$ ($\lesssim200$ MeV) do not greatly alter the 
favorable parameter range for the appearance of quark matter in the
star.  We have also found from the dynamical properties of the 
deconfinement transition that the critical overpressure $\Delta P_{c}$ 
required to form the first droplet in the star narrows the favorable 
parameter range obtained from the static properties into the 
lower $b$ and lower $\alpha_{s}$ regime.  The critical overpressure 
$\Delta P_{c}$, however, is quite uncertain, because it depends on the
poorly known quark-hadron interfacial properties and the unclear 
interactions of quasiparticles in the medium with the droplet surface.
In order to make a 
further investigation, we should take account of the possible effect
of nucleon superfluidity, which acts to decrease the number of
available quasiparticle states with momenta close to the Fermi
surfaces [12].  If the nucleons were in a superfluid state, the 
friction force exerted by the quasiparticles on the droplet and
hence the critical overpressure $\Delta P_{c}$ in the system not 
containing hyperons would be reduced considerably.

It is important to bear in mind that the obtained results for the
deconfinement transition are based on the adopted models for hadronic
matter and for quark matter, which are inequivalent physical 
descriptions of the two phases and hence do not necessarily describe
the transition well.  Granted that such models hold good, the
dynamical properties of the transition are not obvious
%cannot be expressed
quantitatively as mentioned in the preceding paragraph.  Moreover,
the mechanical or chemical impurities leading to the formation
of a quark matter seed,
%transition,
if occurring in the star at all, might relax the condition for
the transition drastically.

Once the first droplet of quark matter arises from fluctuations
in a neutron star, its growth into bulk matter would cause 
strange matter to occupy the central region of the star 
as long as the strange matter is more favorable 
than $\beta$-equilibrated hadronic matter.  These two phases, if both
are present, would have a charged but overall neutral layer which
spreads from the quark-hadron boundary on each side over a length
scale determined by the Thomas-Fermi screening length. At the
boundary, they not only would have the same chemical potential of
leptons, but also would be in equilibrium with respect to $\beta$
and deconfinement processes.  Even if the quark-hadron mixed phase is 
the lowest energy configuration at pressures around the boundary,
its appearance depends on whether the boundary layer is stable against 
fragmentation such as the clustering of protons and of $s$ quarks.
The elucidation of this problem is beyond the scope of this paper.

We believe that the present analysis is applicable to the study of 
the kinetics of the deconfinement transition that may occur in hot
dense matter encountered in stellar collapse and in relativistic 
heavy-ion collisions.  Due to its high temperatures (above $\sim 10$ 
MeV), the nucleation of quark matter in hadronic matter may proceed
classically, i.e., via thermal activation.  Otherwise, thermally
assisted quantum nucleation may take place; it may be influenced by
the irreversible transport of heat and momentum in the medium and by
the dynamical compressibility of the hadronic phase. 
In the case of stellar collapse, neutrinos trapped in a supernova
core, leading to large fractions of electrons and muons, may cause a 
quark matter droplet to have an appreciable electric charge.

%\vspace*{0.4cm}  % {0.5cm}
%\begin{center}
%{\bf ACKNOWLEDGEMENTS}
%\end{center}
%\vspace*{0.3cm}  % {0.5cm} 
\acknowledgments

This work was supported in part by Grants-in-Aid for Scientific 
Research provided by the Ministry of Education, Science, and Culture 
of Japan through Research Grants Nos.\ 07CE2002 and 4396.

%\newpage
%\begin{center}
%{\bf APPENDIX A: DISTRIBUTION INSIDE A QUARK MATTER DROPLET}
%\end{center}
%\vspace*{0.5cm}
\appendix
\section*{DISTRIBUTION INSIDE A QUARK MATTER DROPLET}

A quark matter droplet, if having a macroscopic baryon number and a 
nonzero electric charge in itself, has the constituent quarks 
distributed over a scale of the screening length from the surface just
like a conductor.  In the case of a droplet of $\beta$-stable quark 
matter in vacuum, such a screening action by quarks was investigated
by Heiselberg [40] within the linear Thomas-Fermi approximation [35].
We now extend his calculations to the case of a quark matter droplet 
formed in $\beta$-equilibrated nuclear matter under the flavor 
conservation, considered in Sec.\ IV A 1, and ask to what extent the 
resulting quark distributions alter the potential for the formation
and growth of the droplet.

We begin with the potential energy $U(R)$ of a droplet of radius $R$ 
given by Eq.\ (68), in which we remove the assumption that the quark 
number densities $n_{q}$ ($q=u,d$) are constant inside the droplet,
and we set 
\begin{equation}
%$$ 
n_{q}(r)=\left\{ \begin{array}{lll}
           n_{q0}+\delta n_{q}(r)\ ,
         & \mbox{$r\leq R\ ,$} \\
             \\
           0\ , 
         & \mbox{$r>R\ .$}
 \end{array} \right.
%n_{q}(r)=\theta(R-r)[n_{q0}+\delta n_{q}(r)].
%\eqno{({\rm A}.1)} $$
\end{equation}
Here $n_{u0}$ and $n_{d0}$, independent of $r$, satisfy the pressure 
equilibrium condition (76) and the flavor conservation (61).  In 
determining the deviations $\delta n_{q}$ on a footing equal to 
$\delta n_{l}$, we first note that the baryon density $n_{b,D}$ and
the electric charge density $\rho_{D}$ for the inhomogeneous phase are
written as
\begin{equation}
%$$ 
n_{b,D}(r)=\frac{1}{3}[n_{u}(r)+n_{d}(r)]+\theta(r-R)n_{b,H}
%\eqno{({\rm A}.2)} $$
\end{equation}
and
\begin{eqnarray}
%$$ \hspace{-1cm}  
\rho_{D}(r) &=&
  e\left[\sum_{i=q,l}q_{i}n_{i}(r)+
  \theta(r-R)n_{p}\right]
 \nonumber \\
%$$  
%$$ \hspace*{-1.1cm}  
&=&\theta(R-r)\rho_{Q0}+\delta \rho(r)\ .
%\eqno{({\rm A}.3)} $$
\end{eqnarray}
Here $\delta\rho$ is the deviation of the charge density from that 
calculated in the incompressible limit
\begin{equation}
%$$
\delta \rho(r)=e\left[\theta(R-r)\sum_{q}q_{q}
                 \delta n_{q}+\sum_{l}q_{l}\delta n_{l}\right]
%\eqno{({\rm A}.4)}$$
\end{equation}
%\begin{eqnarray}
% \rho_{D}(r) &=& 
%  e\left[\sum_{i=q,l}q_{i}n_{i}(r)+\theta(r-R)n_{p}\right]
% \nonumber \\
%  &=& \theta(R-r)\rho_{Q0}+e\left[\theta(R-r)\sum_{q}q_{q}
%                  \delta n_{q}+\sum_{l}q_{l}\delta n_{l}\right]\ ,
%\mbox{(A.3)}
%\end{eqnarray}
and $\rho_{Q0}$ is given by Eq.\ (74) in which $n_{u}=n_{u0}$, 
$n_{d}=n_{d0}$, and $n_{s}=0$.  We then expand the potential $U(R)$ up 
to second order in $\delta n_{q}$ and $\delta n_{l}$, and minimize the 
resulting potential with respect to $\delta n_{q}$ and $\delta n_{l}$ 
under the global flavor conservation
\begin{equation}
%$$ 
\int_{r\leq R} dV (f_{0}\delta n_{u}-\delta n_{d})=0\ ,
%\eqno{({\rm A}.5)} $$
\end{equation}
with $f_{0}=n_{d0}/n_{u0}$, and the global neutrality of electric
charge
\begin{equation}
%$$ 
\sum_{l}\int_{V} dV q_{l}\delta n_{l}
+\sum_{q}\int_{r\leq R} dV q_{q}\delta n_{q}+Z_{0}=0\ ,
%\left(\frac{2}{3}\delta n_{u}-\frac{1}{3}
%\delta n_{d}-\delta n_{e}-\delta n_{\mu}\right)+Z_{0}=0\ ,
%\eqno{({\rm A}.6)} $$
\end{equation}
where $Z_{0}$ is given by Eq.\ (73).
This minimization
%, up to $O(\delta n_{i})$,
leads not only to Eq.\ (78) for $l$ leptons, but also to
%and obtain the relations for quarks:
\begin{equation}
%$$ 
\delta n_{q}=-\frac{\kappa_{q}^{2}}{4\pi(q_{q}e)^{2}}
\left[q_{q}e\phi(r)+\mu_{q0}-\frac{1}{3}\mu_{b,H}+q_{q}\mu_{e,H}
+C_{q}\right]
%\eqno{({\rm A}.7)} $$
\end{equation}
for $q$ quarks,
where $\mu_{q0}$ is the chemical potential of $q$ quarks calculated at 
density $n_{q0}$, 
\begin{equation}
%$$
\kappa_{q}=\left[4\pi(q_{q}e)^{2}
\left.\frac{\partial n_{q}}{\partial\mu_{q}}
\right|_{n_{q}=n_{q0}}\right]^{1/2} 
%\eqno{({\rm A}.8)} $$
\end{equation}
is the reciprocal of the Thomas-Fermi screening length of $q$ quarks, 
and $C_{u}=Cf_{0}$ and $C_{d}=-C$ are the constants calculated at a 
given $R$ from the flavor conservation (A5).  Here the proton 
screening continues to be ignored.  We also neglect the energy
increment yielded by the gradient of $\delta n_{q}$ present over a 
scale of the screening length from the droplet surface. 
This increment, acting to flatten the quark distributions, is 
essentially unknown along with the quark-hadron interfacial property.

The electrostatic potential obeys the Poisson equation
\begin{equation}
%$$ 
\nabla^{2}\phi(r)-\kappa(r)^{2}\phi(r)=-4\pi\rho_{Q}\theta(R-r)\ ,
%\eqno{({\rm A}.9)} $$
\end{equation}
where 
\begin{equation}
%$$ 
\rho_{Q}=\rho_{Q0}-
 \sum_{q}\frac{\kappa_{q}^{2}}{4\pi q_{q}e}\left(\mu_{q0}
    -\frac{1}{3}\mu_{b,H}+q_{q}\mu_{e,H}+C_{q}\right)
%\eqno{({\rm A}.10)} $$
\end{equation}
and 
\begin{equation}
%$$ 
\kappa(r)=\left\{ \begin{array}{lll}
      \kappa_{<}\equiv
   \sqrt{\kappa_{L}^{2}+\kappa_{u}^{2}+\kappa_{d}^{2}}\ ,
         & \mbox{$r\leq R\ ,$} \\
                 \\
           \kappa_{L}\ ,
         & \mbox{$r>R\ .$}
 \end{array} \right.
%\kappa_{e}^{2}+\kappa_{\mu}^{2}
%+\theta(R-r)(\kappa_{u}^{2}+\kappa_{d}^{2})]^{1/2}
%\sqrt{\kappa_{L}^{2}+\theta(R-r)\kappa_{Q}^{2}}
%\eqno{({\rm A}.11)} $$
\end{equation}
%with $\kappa_{Q}^{2}=\kappa_{u}^{2}+\kappa_{d}^{2}$.
%is the reciprocal of the screening length of the system.
Here $\kappa_{<}$ ($\kappa_{L}$) corresponds to the reciprocal of the
screening length inside (outside) the droplet.  The solution to 
Eq.\ (A9) is obtained as 
\begin{equation}
%$$   
\phi(r)=\left\{ \begin{array}{ll}
    \displaystyle{\frac{4\pi\rho_{Q}}{\kappa_{<}^{2}}}
    \left[1-
    \displaystyle{\frac{\kappa_{<}R(1+\kappa_{L}R)}
    {\kappa_{L}R\sinh(\kappa_{<}R)+\kappa_{<}R\cosh(\kappa_{<}R)}}
    \displaystyle{\frac{\sinh(\kappa_{<}r)}{\kappa_{<}r}}
         \right]\ ,
         & \mbox{$r\leq R\ ,$} \\
             \\ 
%             \\
    \displaystyle{\frac{4\pi\rho_{Q}}{\kappa_{<}^{2}}}
    \displaystyle{\frac{\kappa_{L}R
     [\kappa_{<}R\cosh(\kappa_{<}R)-\sinh(\kappa_{<}R)]}
     {\kappa_{L}R\sinh(\kappa_{<}R)+\kappa_{<}R\cosh(\kappa_{<}R)}}
    \displaystyle{\frac{\exp[\kappa_{L}(R-r)]}{\kappa_{L}r}}\ ,
         & \mbox{$r>R\ .$}
 \end{array} \right.
%\eqno{({\rm A}.12)} $$
\end{equation}
% $$   \phi(r)=\left\{ \begin{array}{lllll}
%        (4\pi\rho_{Q}/\kappa_{<}^{2})
%        [1-\exp(-\kappa_{<} R)(1+\kappa_{<} R)
%        \sinh(\kappa_{<} r)/\kappa_{<} r]\ ,
%         & \mbox{$r\leq R\ ,$} \\
%             \\ \\
%        (4\pi\rho_{Q}/\kappa_{<}^{2})[\kappa_{<} R\cosh(\kappa_{<} R)
%         -\sinh(\kappa_{<} R)] \\ \\
%        \times\exp[-\kappa_{L} (r-R)-\kappa_{<}R]/\kappa_{<} r\ ,
%         & \mbox{$r>R\ .$}
% \end{array} \right.
%\eqno{({\rm A}.12)} $$
It is straightforward to derive the electrostatic energy $E_{C}$ from 
Eq.\ (A12); the result is
\begin{eqnarray}
%$$ \hspace*{-6cm}
E_{C}(R) &=&
 \displaystyle{\frac{4\pi^{2}\rho_{Q}^{2}}{\kappa_{<}^{5}}}
 \displaystyle{\frac{1}
  {[\kappa_{L}R\sinh(\kappa_{<}R)+\kappa_{<}R\cosh(\kappa_{<}R)]^{2}}} 
\nonumber \\ & &
%$$ $$ 
\times\left\{  \left[
   \kappa_{<}R+\cosh(\kappa_{<}R)\sinh(\kappa_{<}R)
    -\displaystyle{\frac{2\sinh^{2}(\kappa_{<}R)}{\kappa_{<}R}}\right]
     (\kappa_{<}R)^{2}(1+\kappa_{L}R)^{2}\right. 
%$$ $$ 
%\hspace*{-2.8cm}
\nonumber \\ & &
+\left.
    \displaystyle{\frac{\kappa_{<}}{\kappa_{L}}}
    [\kappa_{<}R\cosh(\kappa_{<}R)-\sinh(\kappa_{<}R)]^{2}
    \kappa_{L}R(\kappa_{L}R+2)\right\}\ .
%\eqno{({\rm A}.13)} $$
\end{eqnarray}
%&&   E_{C}(R)=\frac{2\pi^{2}\rho_{Q}^{2}}{\kappa_{<}^{5}}
%  \left\{
%   -3+(\kappa_{<} R)^{2}+\exp(-2\kappa_{<} R)[3+6\kappa_{<} R
%  +5(\kappa_{<} R)^{2}+2(\kappa_{<} R)^{3}] \frac{}{}\right. $$
%$$ \left.
%   +\frac{1}{2}\left(\frac{\kappa_{L}}{\kappa_{<}}-1\right)
%   [\kappa_{<}R-1+(\kappa_{<}R+1)\exp(-2\kappa_{<}R)]^{2}\right\}\ .
%\eqno{({\rm A}.13)} $$
The quark distributions replace the excess lepton energy 
$\Delta E_{L}(R)$ in the potential (68) with the sum of 
$\Delta E_{L}(R)$ and the excess quark energy defined as
\begin{equation}
%$$ 
\Delta E_{Q}(R)=\int_{r\leq R}dV
\{\varepsilon_{Q}[n_{q}(r)]-\varepsilon_{Q}(n_{q0})\}\ .
%\eqno{({\rm A}.14)} $$
\end{equation}
The linear approximation used here yields
\begin{equation}
%$$ 
\Delta E_{L}(R)+\Delta E_{Q}(R)
  =Z_{0}\mu_{e,H}+
   \sum_{q}\left(\mu_{q0}-\frac{1}{3}\mu_{b,H}+q_{q}\mu_{e,H}\right)
   \int_{r\leq R}\delta n_{q}dV\ . 
%\eqno{({\rm A}.15)} $$
\end{equation}

By using Eqs.\ (78), (A4), and (A7), we have calculated the density 
deviations $\delta n_{q}$, $\delta n_{l}$, and $\delta\rho/e$ from the
values obtained not allowing for the quark or lepton screening.
We have confirmed that for $R\lesssim10$ fm, the number density of each 
component remains close to its constant value obtained in the 
incompressible limit.  The results calculated for the inhomogeneous
phase characterized by $b=100$ MeV fm$^{-3}$, $\alpha_{s}=0.4$,
$R=2,5$ fm, and $P=525$ MeV fm$^{-3}$ have been plotted in Fig.\ 9, 
together with the charge density deviation calculated allowing for the
lepton screening alone as in the main text.  We see from the
comparison of these charge density deviations $\delta\rho$ that the 
quark screening does not vary the total charge distribution $\rho_{D}$ 
significantly even for $R=5$ fm, although $\kappa_{<}^{-1}\sim3$ fm is 
%considerably 
small relative to this $R$ and to 
$\kappa_{L}^{-1}\sim5$ fm.  This result reflects the feature that 
under the flavor conservation, the static compressibility of the quark 
fluid plays a role in chemically decreasing the sum of the excess 
lepton and quark energies rather than in reducing the electrostatic 
energy via Coulomb screening.  Note that $\delta n_{q}>0$, leading
to a reduction of the sum $\Delta E_{L}+\Delta E_{Q}$ given by 
Eq.\ (A15).  The resulting decrease in the potential 
$U(R)$ can be observed in Fig.\ 10 by comparing the result obtained
for a lepton- and quark-screened droplet with that for a
lepton-screened droplet.  The case of a lepton- and quark-screened 
droplet agrees fairly well with the case of a lepton-screened droplet 
and the incompressible case in which we set $E_{C}=0$, since the
inhomogeneous phase generally satisfies
$$ Z_{0}\mu_{e,H}\gg Z_{0}\mu_{e,H}-(\Delta E_{L}+\Delta E_{Q})
 > E_{C}\gg|\Delta E_{C}|\ , $$
where $E_{C}$ is given by Eq.\ (A13), and $\Delta E_{C}$ is the 
difference between the right-hand sides of Eqs.\ (80) and (A13).
It is of interest to note that the finite compressibility of the fluid
of $N$ nucleons produces no chemically induced change in their number 
density, as can be found from the expansion of $U(R)$ up to second
order in $\delta n_{N}$.  This property is due to the $\beta$ 
equilibrium sustained in the hadronic phase.

The shifts $\delta n_{u}$ and $\delta n_{d}$ due to the quark 
screening, being typically $\sim0.1$ fm$^{-3}$, in turn affect the 
kinetic properties of the droplet.  First, the effective droplet mass
$M(R)$, estimated from the nucleon kinetic energy as Eq.\ (85), is
increased, since it is proportional to the square of the 
baryon density discontinuity at $r=R$.  Such increase in $M(R)$
diminishes the typical velocity of the precritical droplet,
enforcing its smallness relative to the averaged Fermi velocity 
$v_{F,H}|_{n_{B\neq N}=0}$ given by Eq.\ (89).  On the other hand, the
hydrodynamic mass flow of each quark component, which arises from the 
Coulomb-induced inhomogeneity in $\delta n_{q}$, contributes little
to $M(R)$.  Second, the friction force $F_{H}|_{n_{B\neq N}=0}$ 
given by Eq.\ (87), behaving as $F_{H}|_{n_{B\neq N}=0}\propto M(R)$, 
likewise becomes strong owing to $\delta n_{q}$.  A further friction
force arises from
%; the quark contribution to the friction force, which stems from 
%the gradient of the quark velocity field via 
the collisions of quark quasiparticles, 
carrying the quark mass flow,
%induced by the gradient of the quark velocity field, 
with the droplet surface, but this 
is negligibly small compared with the nucleon contribution.  
Recall that the baryon density discontinuity at
$r=R$ is fairly large, ranging $\sim0.1$--0.7 fm$^{-3}$, in the
absence of the quark screening.  
We may then conclude that the shifts $\delta n_{q}$,
acting to enhance the dissipation effects governed by nucleon 
quasiparticles, lead to a fractional increase in the critical 
overpressure $\Delta P_{c}$.

We conclude this appendix by briefly mentioning the influence of the 
static compressibility of quark matter, surrounded by hadronic matter 
with hyperons, on the potential energy.  This effect can easily be 
estimated by following a line of argument analogous to the case in 
which no hyperons exist and by 
%totally 
neglecting the electrostatic 
properties.  Such an estimate made over the considered range of the 
parameters such as $b$, $\alpha_{s}$, $\sigma_{s}$, and $m_{s}$ 
demonstrates that the finite compressibility increases the number 
densities $n_{u}$, $n_{d}$, and $n_{s}$, independently of $R$,
by less than about 10\% at typical overpressures.  This leads to a 
chemically induced decrease in the quark energy, which has been found 
to be of little consequence to the potential energy.  The resulting 
increase in the baryon density inside a droplet by typically 
$\sim0.05$ fm$^{-3}$ shifts the baryon kinetic energy appreciably,
and hence the effective droplet mass (95).  This is because, in the 
absence of the quark screening, the baryon density discontinuity at 
$r=R$ is relatively small, ranging $\sim -0.1$--0.3 fm$^{-3}$. 
Such shifts move into the lower-$b$ regime the region of the 
bag-model parameters in which the droplet behaves relativistically, 
but otherwise do not change the qualitative
conclusions about the nucleation processes obtained ignoring the quark 
screening.

% now the references. delete or change fake bibitem. delete next three
%   lines and directly read in your .bbl file if you use bibtex.

% \begin{figure}
% \caption{}
% \label{}
% \end{figure}

\begin{figure}
\begin{center}
\leavevmode\psfig{file=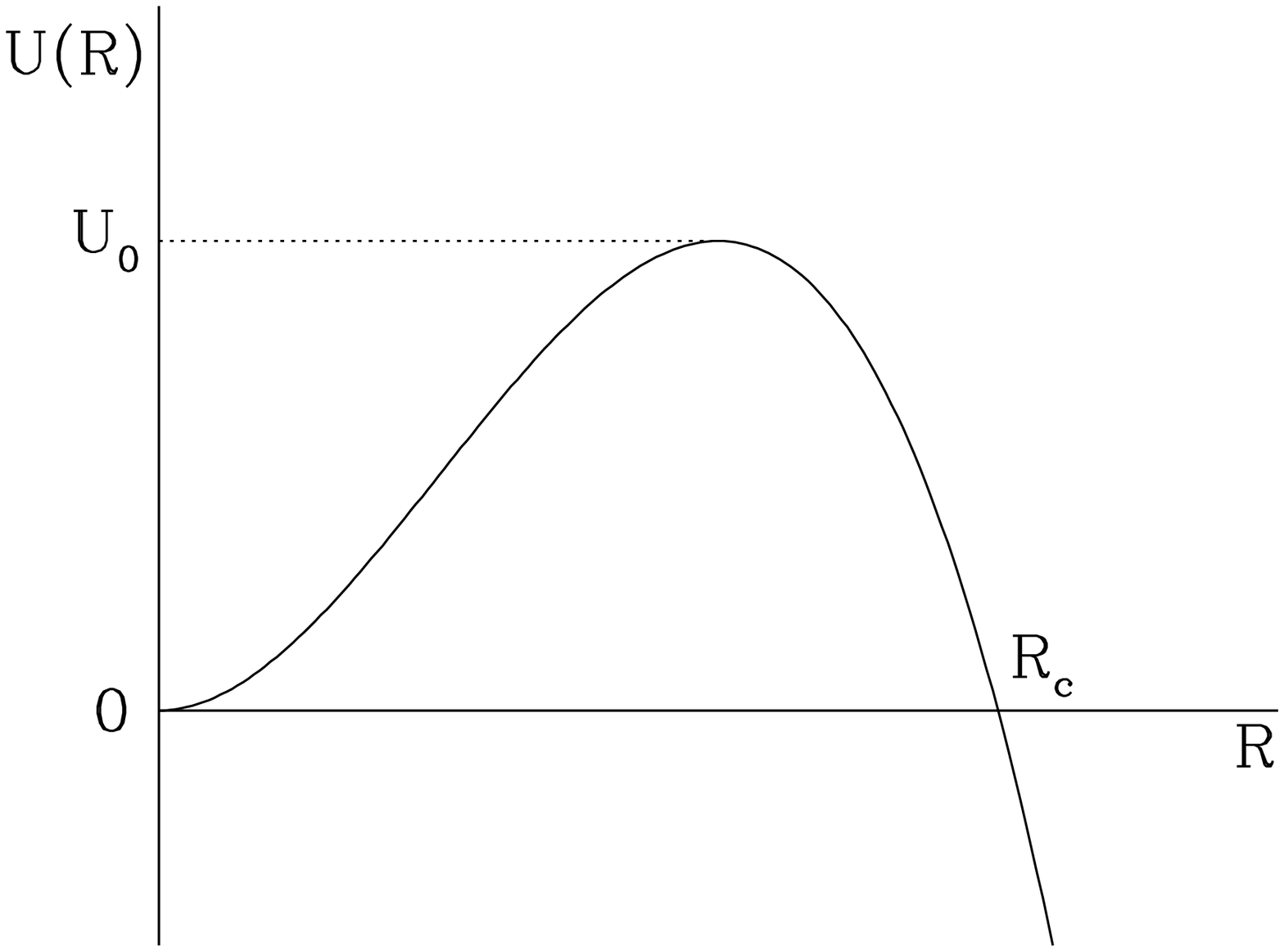,height=15cm}
%\epsfile{file=pot.ps,height=18cm}
\caption{Potential energy for a fluctuation of $R$, the radius of a 
spherical droplet of the stable phase in a single-component system.}
\end{center}
\label{fig1}
\end{figure}

\begin{figure}
\begin{center}
\leavevmode\psfig{file=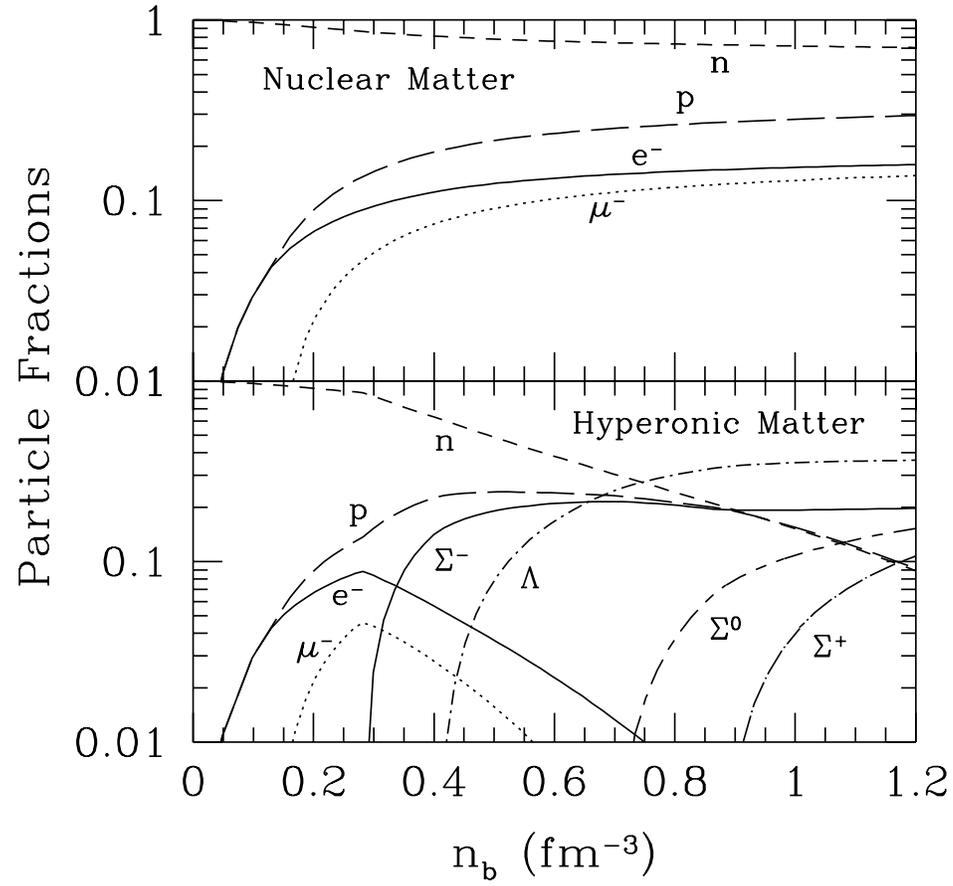,height=15cm}
%\epsfile{file=comrmf.ps,height=18cm}
\caption{Composition of $\beta$-stable nuclear matter (upper panel) 
and hyperonic matter (lower panel) at zero temperature, calculated 
using the model of Glendenning and Moszkowski [16].}
\end{center}
\label{fig2}
\end{figure}

\begin{figure}
\begin{center}
\leavevmode\psfig{file=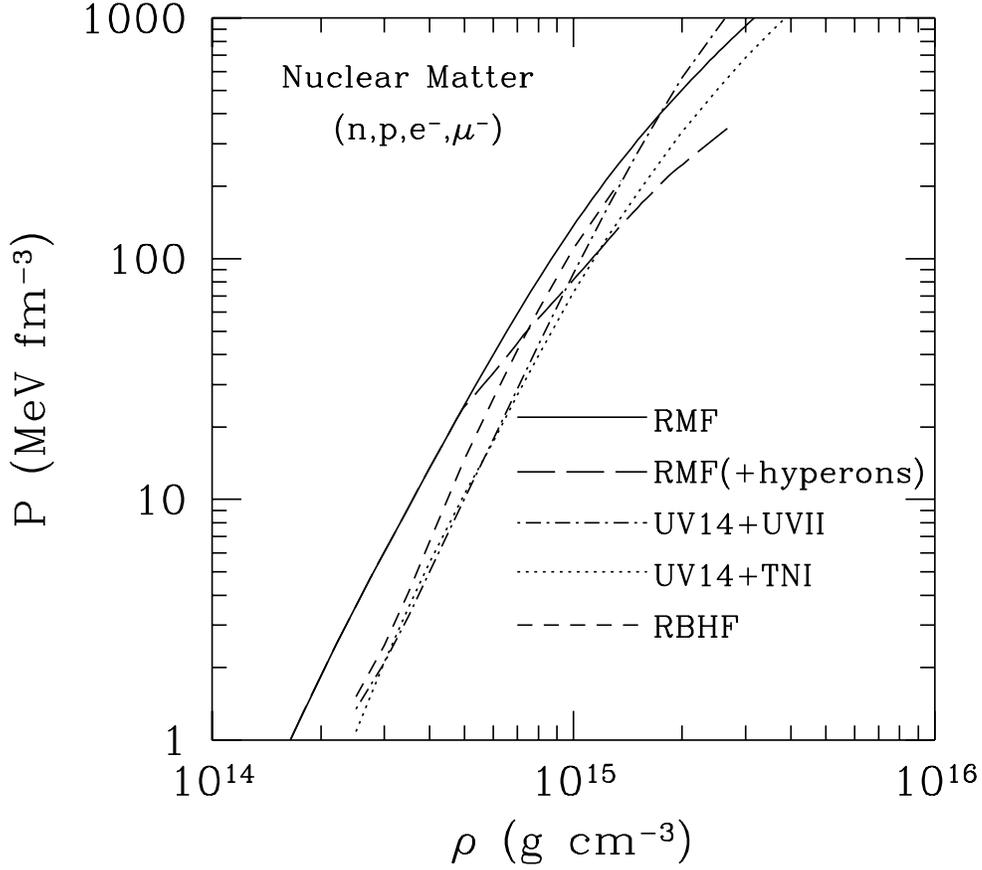,height=15cm}
%\epsfile{file=nmeos2new.ps,height=18cm}
\caption{Equation of state of zero-temperature nuclear matter in 
$\beta$ equilibrium.  The solid and long-dashed lines are the results 
calculated from the model of Glendenning and Moszkowski [16] within
the relativistic mean-field (RMF) theory; the latter has been obtained 
allowing for the presence of hyperons.  The dash-dotted and dotted
lines are the results obtained by Wiringa, Fiks, and Fabrocini [25] 
using the Urbana $v_{14}$ (UV14) two-nucleon potential plus the Urbana
VII (UVII) three-nucleon potential and the density-dependent 
three-nucleon interaction (TNI) model, respectively.  The short-dashed
line is the relativistic Brueckner-Hartree-Fock (RBHF) result
obtained by Engvik $et$ $al$.\ [26] using the Bonn A two-nucleon
potential.}
\end{center}
\label{fig3}
\end{figure}

\begin{figure}
\begin{center}
\leavevmode\psfig{file=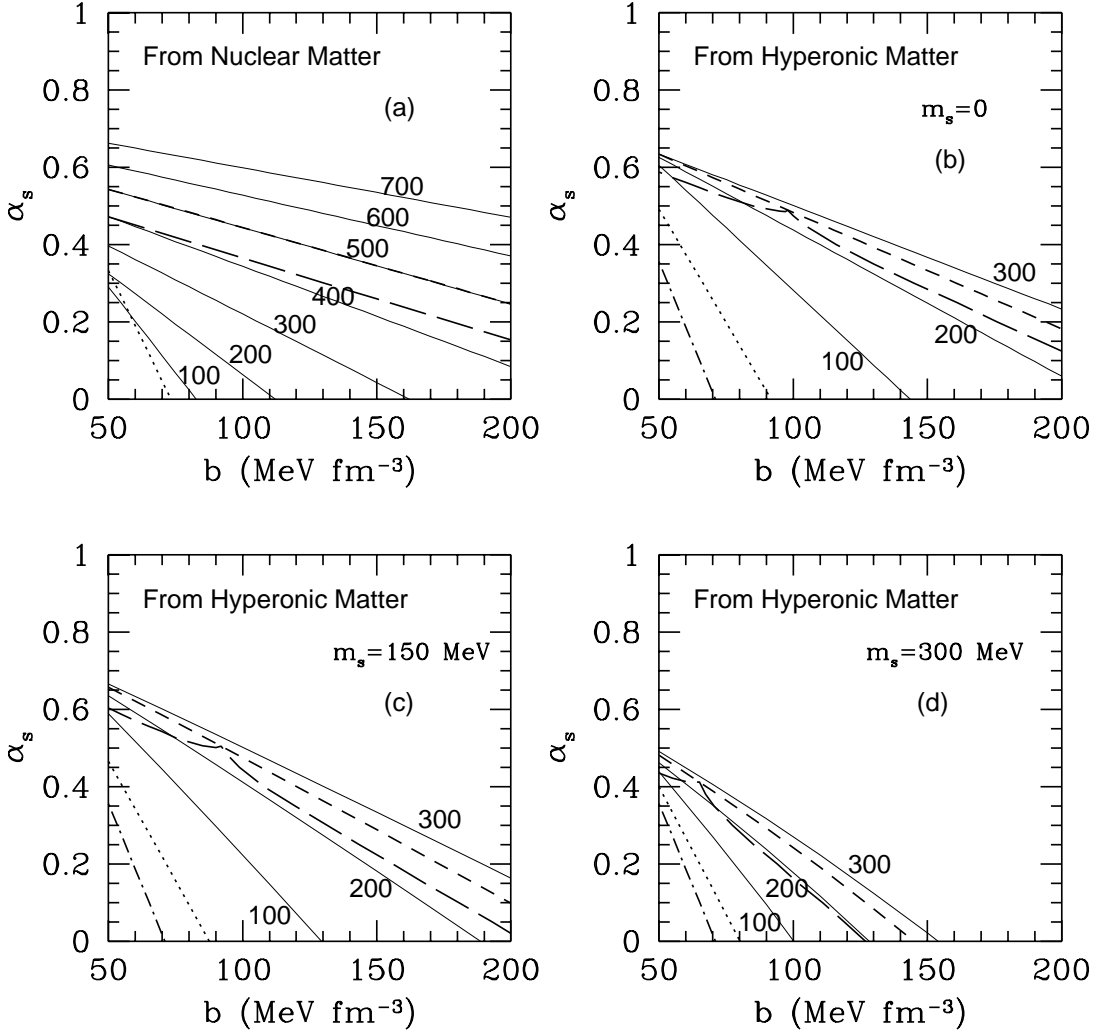,height=15cm}
%\epsfile{file=p0contrmf3.ps,height=18cm}
\caption{Contour plots of the pressure $P_{0}$ (as marked in 
MeV fm$^{-3}$) of the static deconfinement transition on the 
$b$-$\alpha_{s}$ plane.  The solid lines in (a) denote the contours of 
the transition pressures from $\beta$-stable nuclear matter to $u$ and 
$d$ quark matter.  The solid lines in (b), (c), and (d) represent the 
contours of the transition pressures from $\beta$-stable hyperonic
matter to $u$, $d$, and $s$ quark matter calculated for 
$m_{s}=0, 150, 300$ MeV, respectively.  In each panel, the contours of 
$P_{0}=P_{\rm max}$ (short-dashed line), $P_{0}=P_{1.4}$ (dotted
line), and $P_{0}=P_{\Sigma^{-}}$ (dash-dotted line) are included 
(see text).  We have also plotted the combinations of $b$ and 
$\alpha_{s}$ satisfying the critical condition required to form a real 
quark matter droplet via deconfinement in the star that contains the 
nucleation centers of number $N_{s}=10^{48}$ and is spinning down or 
accreting matter from a neighboring star in a time scale of 
$\tau_{s}=1$ Myr.  The lower $b$ and lower $\alpha_{s}$ region is
favorable for the presence of quark matter in the star.  The 
long-dashed line in (a) denotes the result obtained for a droplet 
arising from nuclear matter by setting $\sigma_{s}=30$ MeV fm$^{-2}$ 
and $\alpha=1$; the long-dashed line in (b)--(d), for a droplet 
deconfined from hyperonic matter by setting 
$\sigma_{s}=30$ MeV fm$^{-2}$.}
\end{center}
\label{fig4}
\end{figure}

\begin{figure}
\begin{center}
\leavevmode\psfig{file=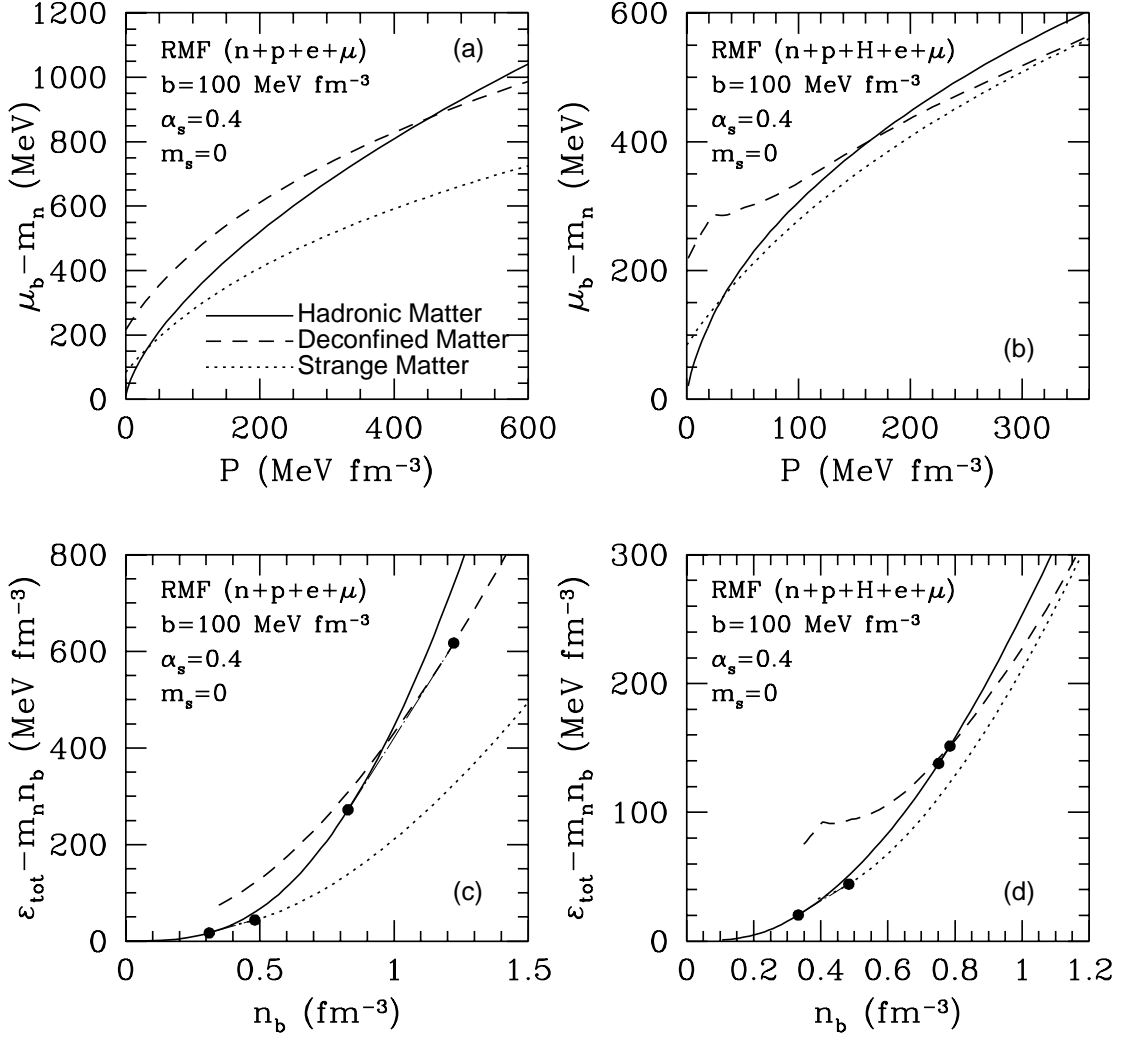,height=15cm}
%\epsfile{file=rmfdia3new2.ps,height=18cm}
\caption{(a) [(b)] Chemical potentials for the electrically neutral
bulk phases as a function of pressure, calculated for the bag-model 
parameters $b=100$ MeV fm$^{-3}$, $\alpha_{s}=0.4$, and $m_{s}=0$.  
The solid line represents the phase of $\beta$-equilibrated hadronic 
matter without (with) hyperons; the dashed line, the phase of matter 
deconfined therefrom; the dotted line, the phase of strange matter, 
i.e., $u$, $d$, and $s$ quark matter in $\beta$ equilibrium.
(c) [(d)] Energy densities for the corresponding phases as a function
of baryon density.  The dash-dotted lines are the double-tangent 
constructions denoting the coexistence of the two bulk phases
involved.}
\end{center}
\label{fig5}
\end{figure}

\begin{figure}
\begin{center}
\leavevmode\psfig{file=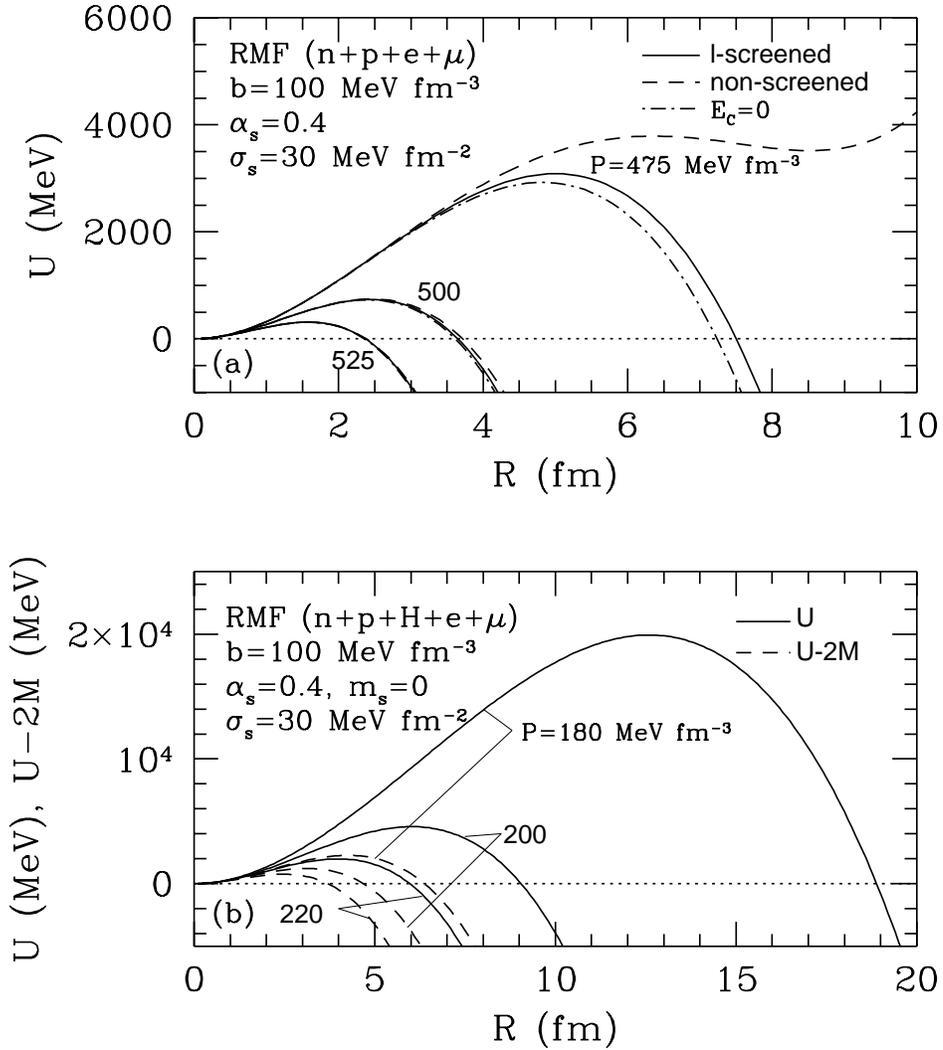,height=15cm}
%\epsfile{file=rmfpot4.ps,height=18cm}
\caption{Potential energies $U(R)$ of a quark matter droplet present
in $\beta$-stable nuclear matter (a) and hyperonic matter (b), 
evaluated for the bag-model parameters $b=100$ MeV fm$^{-3}$, 
$\alpha_{s}=0.4$, and $m_{s}=0$ and for various pressures above the 
pressure $P_{0}$ of the static deconfinement transition.  In (a), we 
have set $\sigma_{s}=30$ MeV fm$^{-2}$.  The solid lines are the
results obtained from Eq.\ (68) for a lepton-screened droplet; the 
dashed lines, for a nonscreened droplet; the dash-dotted lines, in
the case in which we set $E_{C}=0$.  The solid lines in (b) are the 
results evaluated from Eq.\ (92) for $\sigma_{s}=30$ MeV fm$^{-2}$. 
The quantities $U(R)-2M(R)$, where $M(R)$ is the effective droplet
mass given by Eq.\ (95), have also been plotted in (b) by dashed
lines.  The maximum value of $U(R)-2M(R)$ yields the upper bound of
the forbidden region of the droplet energy.}
\end{center}
\label{fig6}
\end{figure}

\begin{figure}
\begin{center}
\leavevmode\psfig{file=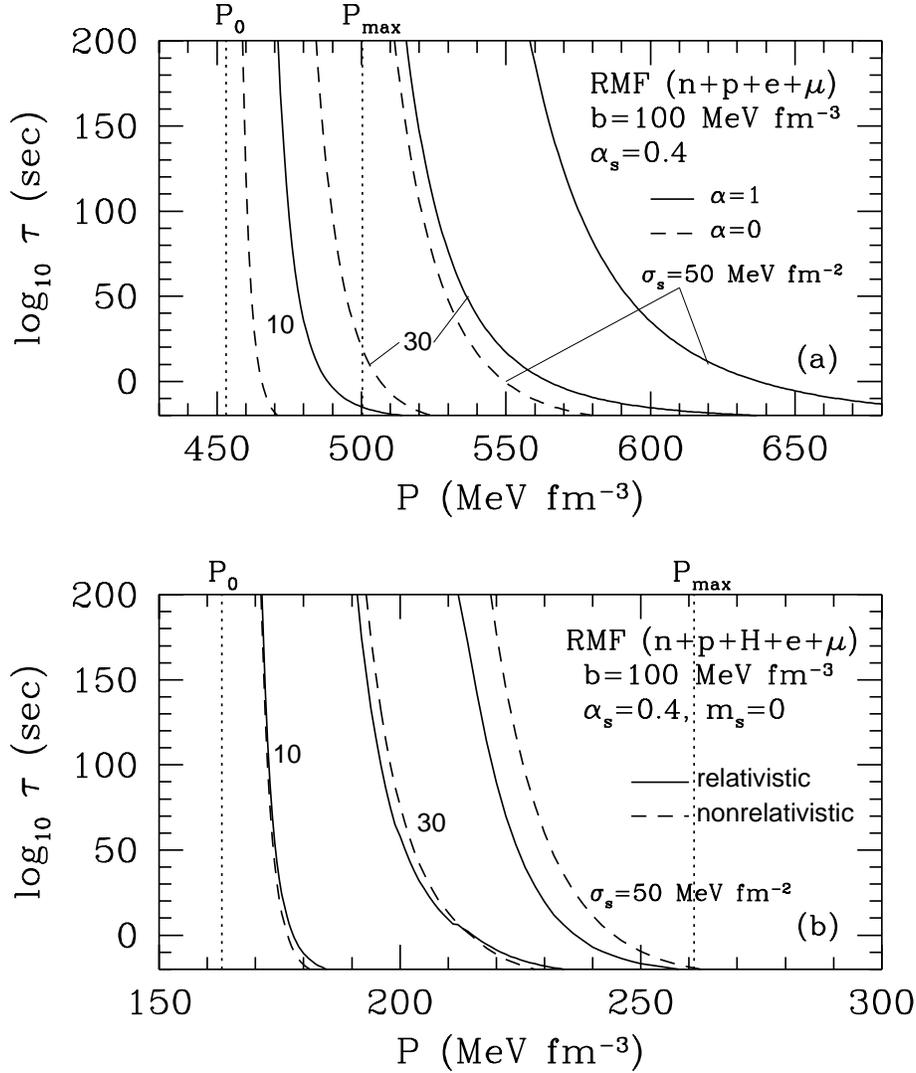,height=15cm}
%\epsfile{file=rmfwkb4.ps,height=18cm}
\caption{Time needed to form a real quark matter droplet via quantum
tunneling in $\beta$-stable zero-temperature nuclear matter (a) and 
hyperonic matter (b), evaluated as a function of pressure for the 
bag-model parameters $b=100$ MeV fm$^{-3}$, $\alpha_{s}=0.4$, and 
$m_{s}=0$.  In (a), the surface tension has been set as
$\sigma_{s}=10,30,50$ MeV fm$^{-2}$.
The solid lines are the dissipative results obtained for $\alpha=1$, 
and the dashed lines are the nondissipative results.  In (b), the 
relativistic (solid lines) and nonrelativistic (dashed lines) results 
calculated for $\sigma_{s}=10,30,50$ MeV fm$^{-2}$ are plotted. 
For comparison, we have marked in both panels
the pressure $P_{0}$ of the static 
deconfinement transition and the central pressure $P_{\rm max}$ of the
maximum-mass neutron star.}
\end{center}
\label{fig7}
\end{figure}

\begin{figure}
\begin{center}
\leavevmode\psfig{file=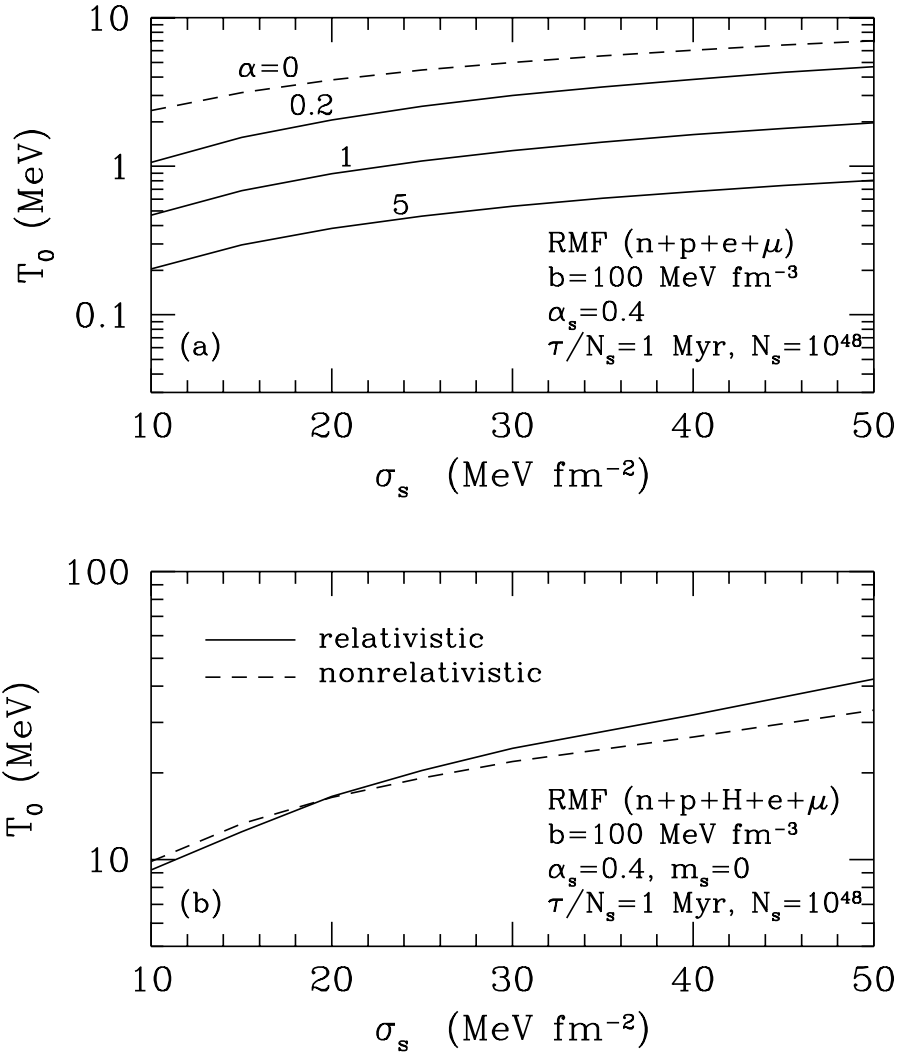,height=15cm}
%\epsfile{file=rmft03.ps,height=18cm}
\caption{Crossover temperatures $T_{0}$ from the quantum-tunneling
to the thermal-activation nucleation of quark matter in $\beta$-stable
nuclear matter (a) and hyperonic matter (b), calculated as a function
of the surface tension for the bag-model parameters 
$b=100$ MeV fm$^{-3}$, $\alpha_{s}=0.4$, and $m_{s}=0$ and for the 
stellar conditions $\tau/N_{s}=1$ Myr and $N_{s}=10^{48}$.  In (a),
the solid lines are the results obtained for dissipative cases of 
$\alpha=0.2, 1, 5$, and the dashed line is the nondissipative result.
In (b), the solid and dashed lines are the relativistic and 
nonrelativistic results, respectively, estimated not allowing for 
thermal effects on the quantum nucleation.}
\end{center}
\label{fig8}
\end{figure}

\begin{figure}
\begin{center}
\leavevmode\psfig{file=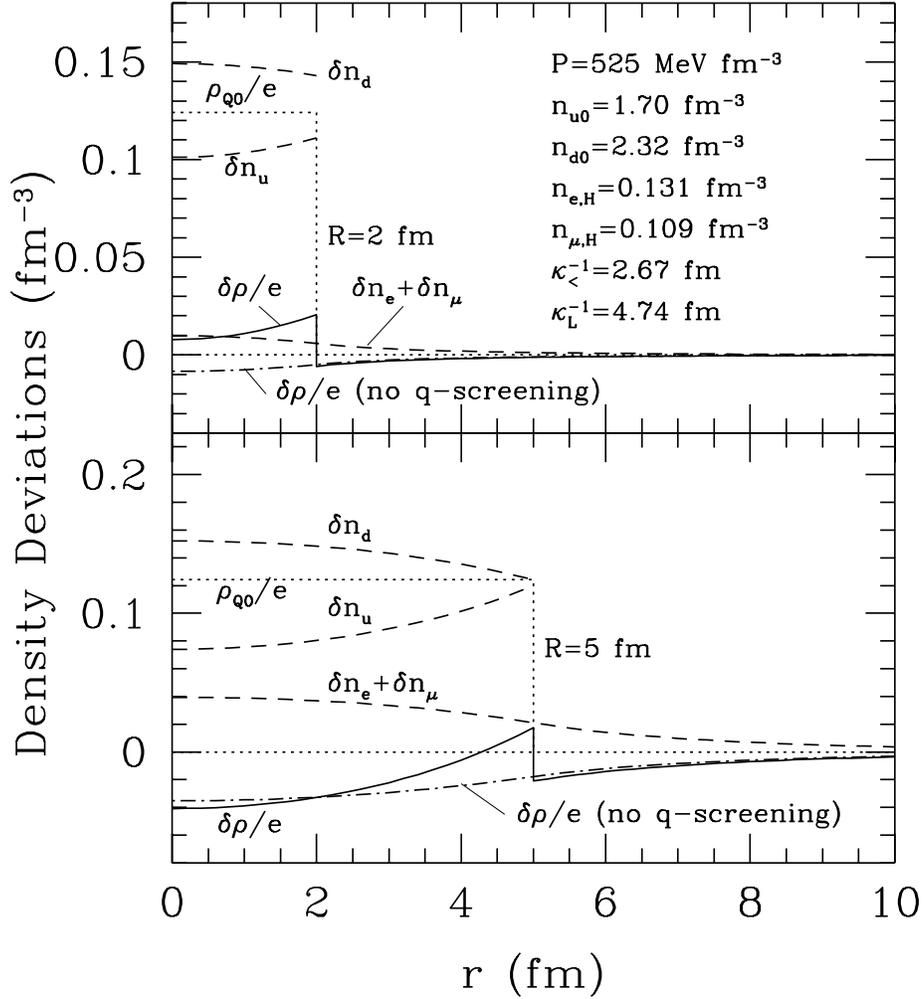,height=15cm}
%\epsfile{file=qscdnnew.ps,height=18cm}
\caption{Deviations $\delta n_{u}$, $\delta n_{d}$, and 
$\delta n_{e}+\delta n_{\mu}$ of the quark and lepton number densities
from the values $n_{u0}$, $n_{d0}$, and $n_{e,H}+n_{\mu,H}$ obtained
in the incompressible limit due to the screening (static 
compressibility) of the fluids of quarks and leptons, calculated 
ignoring the presence of hyperons for the droplet radius $R=2$ fm 
(upper panel) and $R=5$ fm (lower panel), the pressure 
$P=525$ MeV fm$^{-3}$, and the bag-model parameters 
$b=100$ MeV fm$^{-3}$ and $\alpha_{s}=0.4$.  The resulting deviation
$\delta\rho$ of the electric charge density from the stepwise charge 
distribution obtained in the incompressible limit has also been
plotted, together with the result allowing for the lepton screening 
alone (no quark screening).  Since $\delta n_{e}$ graphically agrees
with $\delta n_{\mu}$, only the sum $\delta n_{e}+\delta n_{\mu}$ has
been shown.  See the text for the definition of the screening lengths
$\kappa_{<}^{-1}$ and $\kappa_{L}^{-1}$.}
\end{center}
\label{fig9}
\end{figure}

\begin{figure}
\begin{center}
\leavevmode\psfig{file=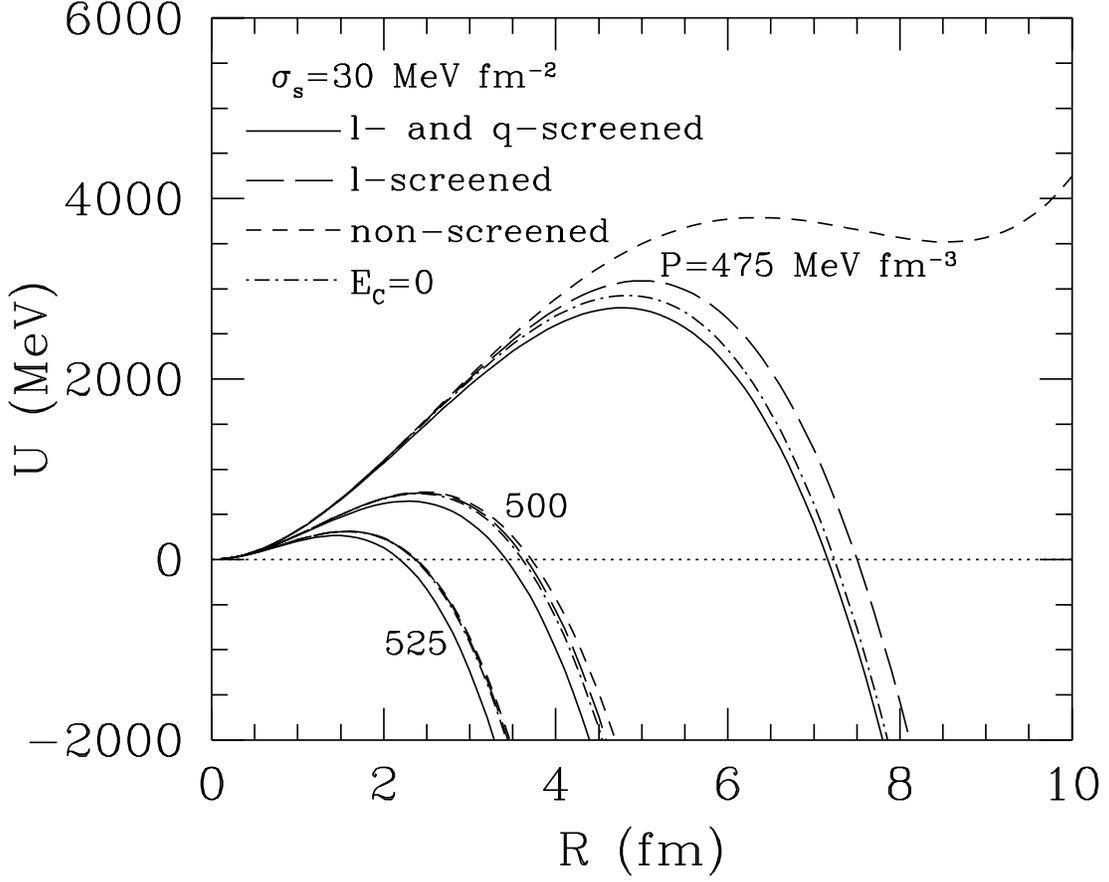,height=15cm}
%\epsfile{file=qscpotnew2.ps,height=18cm}
\caption{Potential energies $U(R)$ of a quark matter droplet in the
metastable phase of $\beta$-stable nuclear matter evaluated for the 
surface tension $\sigma_{s}=30$ MeV fm$^{-2}$, the bag-model
parameters $b=100$ MeV fm$^{-3}$ and $\alpha_{s}=0.4$, and various 
pressures above the pressure, $P_{0}\approx 453$ MeV fm$^{-3}$, of the
static deconfinement transition.  The solid lines are the results 
obtained from Eq.\ (68) for a lepton- and quark-screened droplet; the
long-dashed lines, for a lepton-screened droplet; the short-dashed
lines, for a nonscreened droplet; the dash-dotted lines, in the 
incompressible case in which we set $E_{C}=0$.}
\end{center}
\label{fig10}
\end{figure}

\newpage

% \begin{table}
% \caption{}
% \label{}
% \end{tabular}
% \end{table}

\begin{table}
\caption{Values of $M_{\rm max}$, $P_{\rm max}$, $P_{1.4}$, 
and $P_{\Sigma^{-}}$ evaluated for hyperonic and nuclear matter using
the relativistic mean-field theory with the parameters taken from
Ref.\ [16].}
\label{table1}
\begin{tabular}{ccccc} 
%\hline \hline
   & & $P_{\rm max}$ &$P_{1.4}$ & $P_{\Sigma^{-}}$ \\ 
   & $M_{\rm max}$ & (MeV fm$^{-3}$) & (MeV fm$^{-3}$)& (MeV fm$^{-3}$)
\\ \hline 
Hyperonic matter & 1.78$M_{\odot}$ & 261.0 & 43.13 & 23.30 \\
Nuclear matter & 2.36$M_{\odot}$ & 500.4 & 38.91 & 
%\\ \hline \hline
\end{tabular}
\end{table}

\end{document}